\documentclass[floatfix,aps,twocolumn,superscriptaddress,pra,altaffillsymbol]{revtex4-1}

\usepackage{graphicx}
\usepackage{amsmath}
\usepackage{natbib}
\usepackage{chngpage}
\usepackage{color}

\usepackage{dcolumn}
\usepackage{bm}
\usepackage{hyperref}
\usepackage{amssymb}
\usepackage{float}
\usepackage{epsfig}
\usepackage{blindtext}
\usepackage[utf8]{inputenc}
\usepackage{silence}
\newcommand{\BibitemShut}[1]{}
\makeatletter
\renewcommand{\p@subsection}{}
\renewcommand{\p@subsubsection}{}
\makeatother

\begin{document}

\title{The role of the nominal iron content in the structural, compositional and physical properties of BaFe$_{2+\delta}$S$_3$}

\author{M. L. Amig\'o}
\affiliation{Leibniz IFW Dresden, Helmholtzstr. 20, 01069 Dresden, Germany}
\author{Q. Stahl}
\affiliation{Institut f\"ur Festk\"orper- und Materialphysik, Technische Universität Dresden, 01069 Dresden, Germany}
\author{A. Maljuk}
\affiliation{Leibniz IFW Dresden, Helmholtzstr. 20, 01069 Dresden, Germany}
\author{A. U. B. Wolter}
\affiliation{Leibniz IFW Dresden, Helmholtzstr. 20, 01069 Dresden, Germany}
\author{C. Hess}\altaffiliation[Present address: ]{Fakult\"at f\"ur Mathematik und Naturwissenschaften, Bergische Universit\"at Wuppertal, 42097 Wuppertal, Germany}
\affiliation{Leibniz IFW Dresden, Helmholtzstr. 20, 01069 Dresden, Germany}
\author{J. Geck}
\affiliation{Institut f\"ur Festk\"orper- und Materialphysik, Technische Universität Dresden, 01069 Dresden, Germany}
\affiliation{W\"urzburg-Dresden Cluster of Excellence ct.qmat, Technische Universit\"at Dresden, 01062 Dresden, Germany}
\author{S. Wurmehl}
\affiliation{Leibniz IFW Dresden, Helmholtzstr. 20, 01069 Dresden, Germany}
\author{S. Seiro}
\affiliation{Leibniz IFW Dresden, Helmholtzstr. 20, 01069 Dresden, Germany}
\author{B. B\"uchner}
\affiliation{Leibniz IFW Dresden, Helmholtzstr. 20, 01069 Dresden, Germany}
\affiliation{Institut f\"ur Festk\"orper- und Materialphysik, Technische Universität Dresden, 01069 Dresden, Germany}
\affiliation{W\"urzburg-Dresden Cluster of Excellence ct.qmat, Technische Universit\"at Dresden, 01062 Dresden, Germany}

\date{\today}

\begin{abstract}

BaFe$_2$S$_3$ is a quasi-one-dimensional antiferromagnetic insulator that becomes superconducting under hydrostatic pressure. The magnetic ordering temperature, $T_N$, as well as the presence of superconductivity have been found to be sample dependent. It has been argued that the Fe content may play a decisive role, with the use of 5\%mol excess Fe being reportedly required during the synthesis to optimize the magnetic ordering temperature and the superconducting properties. However, it is yet unclear whether an Fe off-stoichiometry is actually present in the samples, and how it affects the structural, magnetic and transport properties. Here, we present a systematic study of compositional, structural and physical properties of BaFe$_{2+\delta}$S$_3$ as a function of the nominal Fe excess $\delta$. As $\delta$ increases, we observe the presence of an increasing fraction of secondary phases but no systematic change in the composition or crystal structure of the main phase. Magnetic susceptibility curves are influenced by the presence of magnetic secondary phases. The previously reported maximum of $T_N$ at $\delta$=0.1 was not confirmed. 
Samples with nominal $\delta$=0 present the lowest $T_N$ and the resistivity anomaly at the highest temperature $T^*$ while, for $\delta \geq 0.05$, both quantities and the transport gap are seemingly $\delta$-independent. 
Finally, we show that crystals free of ferromagnetic spurious phases can be obtained by remelting samples with nominal $\delta$=0.05 in a Bridgman process.

\end{abstract}

\maketitle

\section{Introduction}\label{Introduction}

In 2015, superconductivity under hydrostatic pressure was discovered in BaFe$_2$S$_3$ \cite{takahashi2015pressure} and two years later in the related compound BaFe$_2$Se$_3$ under similar conditions \cite{ying2017interplay}. Both chalcogenides present two salient differences to previously known iron-based superconductors. First, these materials are insulators and only become metallic and superconducting under pressure \cite{takahashi2015pressure, yamauchi2015pressure}. Second, they do not present a two-dimensional layered crystalline structure, but rather
a quasi-one dimensional structure consisting of FeS$_4$ tetrahedra forming two-leg ladders separated by Ba atoms \cite{hong1972crystal}.
The interest in materials having this geometry can be traced back to copper oxide structures. 
Dagotto and Rice proposed that quasi-one-dimensional cuprate quantum magnets already exhibit some of the key properties of the layered high-T$_\text{c}$ cuprates, such as a spin gap, reminiscent of the pseudogap phase, or the emergence of superconductivity upon hole doping \cite{dagotto1996surprises}. 
Two years later, these predictions were materialized in the cuprate Sr$_{14-x}$Ca$_x$Cu$_{24}$O$_{41}$ \cite{magishi1998spin}. 
These qualitative aspects make the iron-based BaFe$_2$(S,Se)$_3$ family a novel interesting platform to revisit the interplay between electronic correlations, effective dimensionality and superconductivity \cite{craco2018microscopic, roh2020magnetic, PhysRevMaterials3014801, PhysRevB95115154}.

At ambient pressure, BaFe$_2$S$_3$ shows a stripe-type antiferromagnetic order below $T_N\sim$120\,K \cite{takahashi2015pressure, yamauchi2015pressure, materne2018bandwidth1}. As a function of hydrostatic pressure ($P$), $T_N$ first increases and then presents an abrupt reduction at pressures near the insulator to metal transition ($P\sim$10\,GPa) \cite{materne2018bandwidth1, zheng2018gradual}. 
In the metallic phase, superconductivity emerges, reaching a maximum critical temperature of $\sim$24\,K \cite{takahashi2015pressure, yamauchi2015pressure}. Both, the presence of antiferromagnetic order and the emergence of superconductivity under applied pressure are similar to those of the parent compounds of the 122 (for example CaFe$_2$As$_2$ \cite{torikachvili2008pressure}) and 1111 (for example LaFeAsO \cite{okada2008superconductivity}) iron pnictide families.

However, the reported properties of BaFe$_2$S$_3$ samples depend on the synthesis procedure \cite{takahashi2015pressure, PhysRevB.101.205129}. In Ref. \cite{takahashi2015pressure}, samples grown using excess Fe were found to exhibit the largest $T_N$, as well as superconductivity under hydrostatic pressure, while samples grown from a stoichiometric ratio of precursors yielded a low $T_N$ and no superconductivity up to 20\,GPa. This suggested the presence of a slight Fe deficiency as a detrimental factor for both magnetic order and superconductivity.  A systematic study as a function of the excess Fe used in the synthesis reported the maximal $T_N$, identified as a dip in the magnetic susceptibility, for samples grown using 5\%mol excess Fe. The samples grown starting from this composition were assumed to present the true 123 stoichiometry based on energy-dispersive x-ray spectroscopy and resistivity measurements \cite{hirata2015effects}.

The correlation of the actual Fe content in the crystals with the crystal structure and the characteristic temperatures is still unknown, mostly due to the paucity of systematic studies.
It is yet unclear whether the excess Fe used in the reaction is really incorporated into the crystal structure or only forms spurious phases and, in the last case, which is the effect of these extra phases in the physical properties like magnetization or resistivity. 
In this work, we fill this gap presenting a comprehensive study of BaFe$_{2+\delta}$S$_3$ as a function of the nominal Fe excess, $\delta$, used during growth. 
For this, we characterized in detail the composition and the crystal structure as a function of $\delta$. In the light of this study, we present resistivity and magnetization measurements to track the evolution of the N\'eel temperature and the effect of spurious phases.
This article is organized as follows. In Section \ref{methods}, methodological aspects are presented. In Section \ref{section-comp}, the topography and the composition of our samples are discussed. Section \ref{section:rx} is devoted to powder and single crystal x-ray diffraction results. Sections \ref{magnetization} and \ref{resistivity} discuss magnetization and resistivity results, respectively, and present the evolution of $T_N$ with $\delta$. Finally, Section \ref{conclusions} contains our conclusions.

\section{Methods}
\label{methods}

Single crystals were grown from powders of BaS (Alfa Aesar 99.7\%), Fe (Acros organics 99\%) and S (Alfa Aesar 99.5\%) in the molar ratio 1:2+$\delta$:2 with $\delta$=0, 0.05, 0.1 and 0.2. The powders were thoroughly mixed using an agate mortar inside a glovebox under an inert Ar atmosphere and placed in a carbon glass crucible. A quartz ampule containing the crucible was sealed under vacuum. We placed the quartz ampule in a vertical position inside a Nabertherm programmable chamber furnace and heated up to 1100\,$^\circ$C. This temperature is above the reported melting points of BaFe$_2$S$_3$ \cite{hong1972crystal} but below the melting point of BaS \cite{stinn2017experimentally} and Fe \cite{SpringerMaterialsAdditionalResources2010:sm_nlb_978-3-540-77877-6_26}. Then, the temperature was lowered to 750\,$^\circ$C at 3\,$^\circ$C/h. From this slow decrease through the melting point of BaFe$_2$S$_3$, we obtained an ingot at the bottom of the crucible from which needle-like single crystals of mm length could be mechanically detached. 

For the case of $\delta$=0.05, we also studied the effect of re-melting the obtained crystals using a Bridgman procedure in a floating-zone furnace.
For this, the ingot obtained in the previous step was ground into a powder and put in a quartz ampule. 
The powder was melted by optical heating in a 4-mirror type image furnace produced by CSI (Japan) using 1\,kW halogen lamps as a heat source. The completeness of powder melting was controlled by means of a CCD camera and direct visual (by protected eyes) observation. It is worth to mention that optical heating provides a steeper temperature gradient on the crystallization front compared to a conventional Bridgman method based on resistive heating. Prior to the growth, the furnace chamber was evacuated up to 0.01\,mbar and purged with Ar (5N) gas 3 times in order to remove oxygen from the chamber atmosphere. The quartz container had an inner diameter of 11\,mm and wall thickness of about 1.5\,mm. Quartz glass presents the advantage of an extremely high thermal shock resistance, good optical transmission in visible and near-IR ranges and is relatively chemically inert. The initial powder was melted at 73\% of lamps power under 7.5\,bars Ar pressure. Then the glass container with the melt was pulled down at 1.6\,mm/h and it was rotated at 12\,rpm. A solidified ingot of $\sim$15\,mm in length was mechanically detached from the cracked quartz container, from which single crystals could be easily broken off. Only slight traces of evaporation were observed on the inner wall of the quartz ampule.

In the rest of this work, we are going to identify the samples by the nominal Fe excess, $\delta$, and use the label ``re-melted" for the samples grown using the Bridgman technique.

We studied the topography and the composition of the samples in a Zeiss EVOMA15 scanning electron microscope (SEM) with AzTec software equipped with an electron microprobe analyzer for semi-quantitative elemental analysis using the energy dispersive x-ray (EDX) mode. To improve comparability of results between samples, we used a common plane surface. For this, we embedded the crystals together prior to polishing. We used  BaF$_2$ and FeS$_2$ as standards.
We performed more than 50 measurements for each batch of samples, retaining only data sets where the total weight percent (Wt\%) was between 95 and 105 Wt\%.

Powder x-ray diffraction experiments were carried out at room temperature on crushed single crystals in a STOE STADI P diffractometer equipped with a MYTHEN 2K detector using  Mo-\(K_{\alpha 1}\) radiation.
The data were analyzed using the FullProf Suite program \cite{rodriguez1993recent}. The March-Dollase multi-axial model for preferred orientation was used in the refinements.

Single-crystal x-ray diffraction data were collected at 295 K on a Bruker-AXS KAPPA APEX II CCD diffractometer with graphite-monochromated Mo-\(K_{\alpha}\) x-ray radiation (50 kV, 30 mA). The crystal-to-detector distance was 45.1 mm and the detector was positioned at a \(2{\Theta}\) position of 30$^\circ$ for the measurements using  an \({\omega}\)-scan mode strategy at four different \({\phi}\) positions (0$^\circ$, 90$^\circ$, 180$^\circ$ and 270$^\circ$). All data processing was performed in the Bruker APEX3 software suite \cite{APEX3}, the reflection intensities were integrated using SAINT \cite{SAINT} and multi-scan absorption correction was applied using SADABS \cite{SADABS}. The subsequent weighted full matrix least-squares refinements on F$^{2}$ were done with SHELX-2012 \cite{SHELX} as implemented in the WinGx 2014.1 program suite \cite{WinGX}. The crystal structures were refined with anisotropic displacement parameters for all atoms. The used lattice constants were determined via Rietveld refinement of the corresponding powder patterns.

Magnetization measurements were carried out in a Vibrating Sample Superconducting Quantum Interference Device Magnetometer (SQUID-VSM) from Quantum Design. The magnetic field was applied parallel to the needle direction. For all measurements, we used a zero field cooling procedure (ZFC). To reduce the value of the residual magnetic field, to less than 2\,Oe, 
we applied a field of 5\,T at room temperature and then removed it in an oscillation mode.  This procedure was performed before each temperature or field dependence magnetization measurement.

For the resistivity measurements, we used a dip-stick setup immersed in a Dewar with liquid He. A standard four probe method was used. The electric contacts were made with silver paste and gold wires of a diameter $\sim$25\,$\mu$m. The current was applied along the needle direction.

\section{Results}

\subsection{Topography and composition characterization}
\label{section-comp}

\begin{figure}[h!]
\begin{center}
\includegraphics[width=0.22\textwidth]{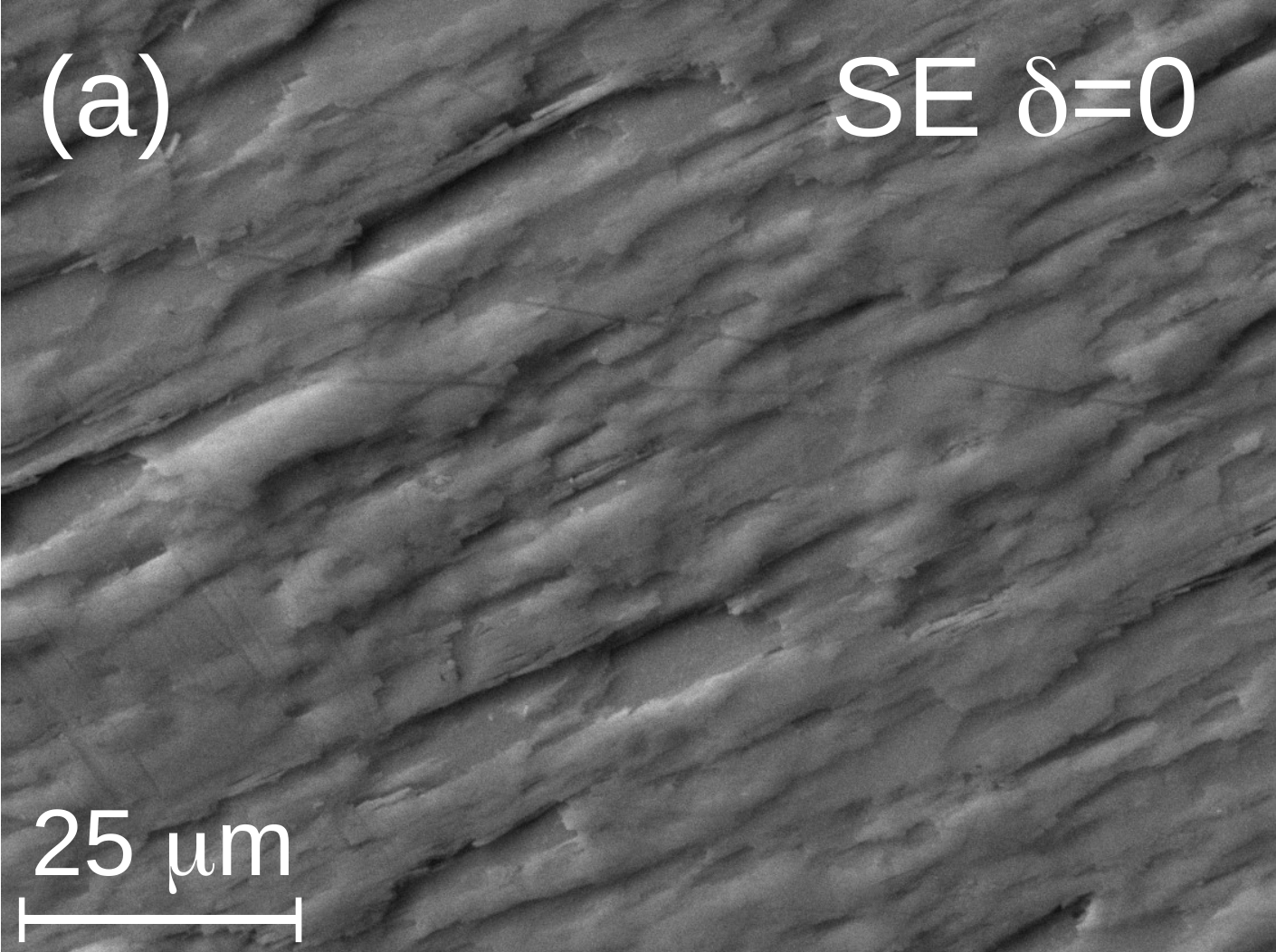}
\includegraphics[width=0.22\textwidth]{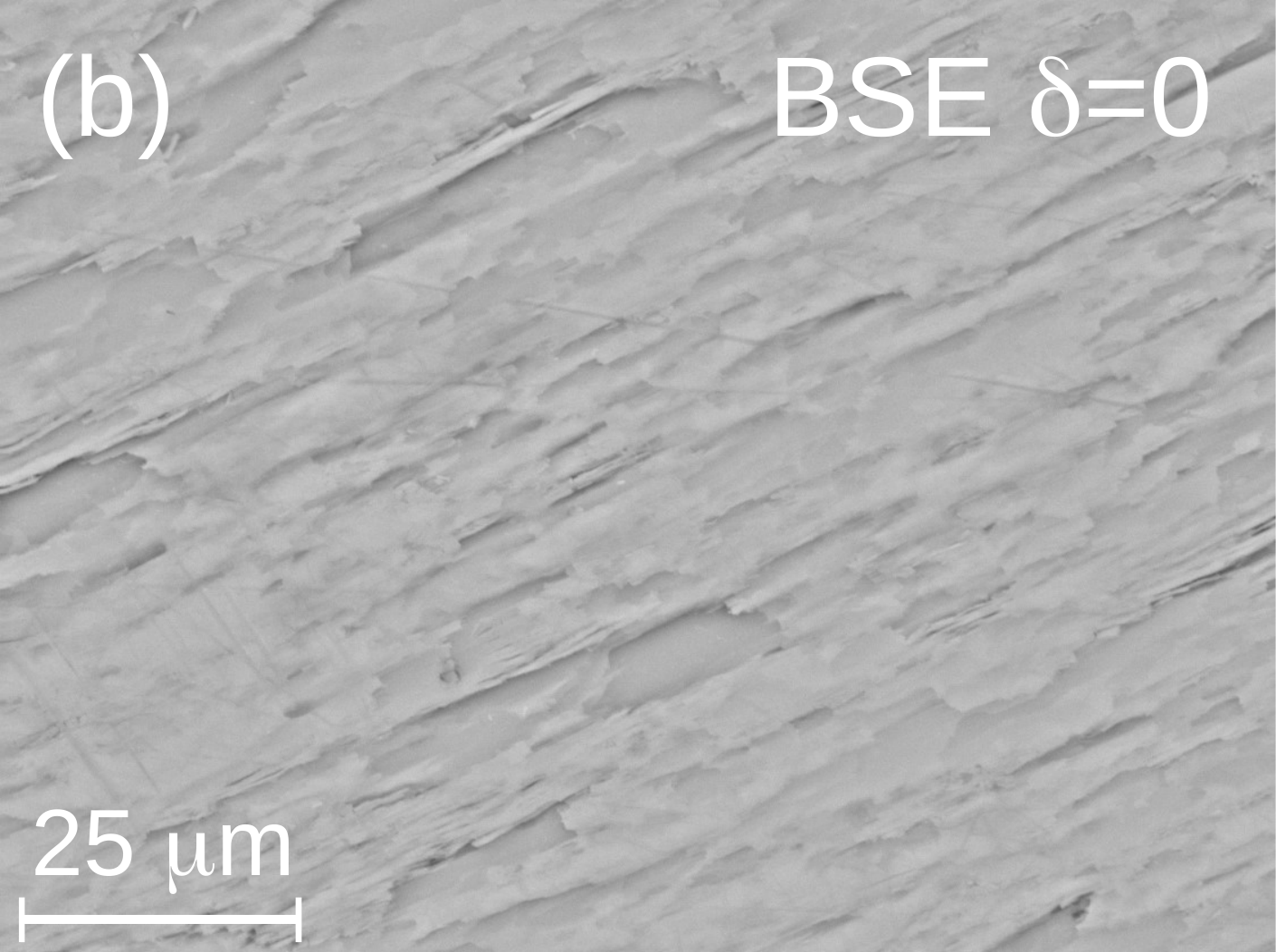}
\includegraphics[width=0.22\textwidth]{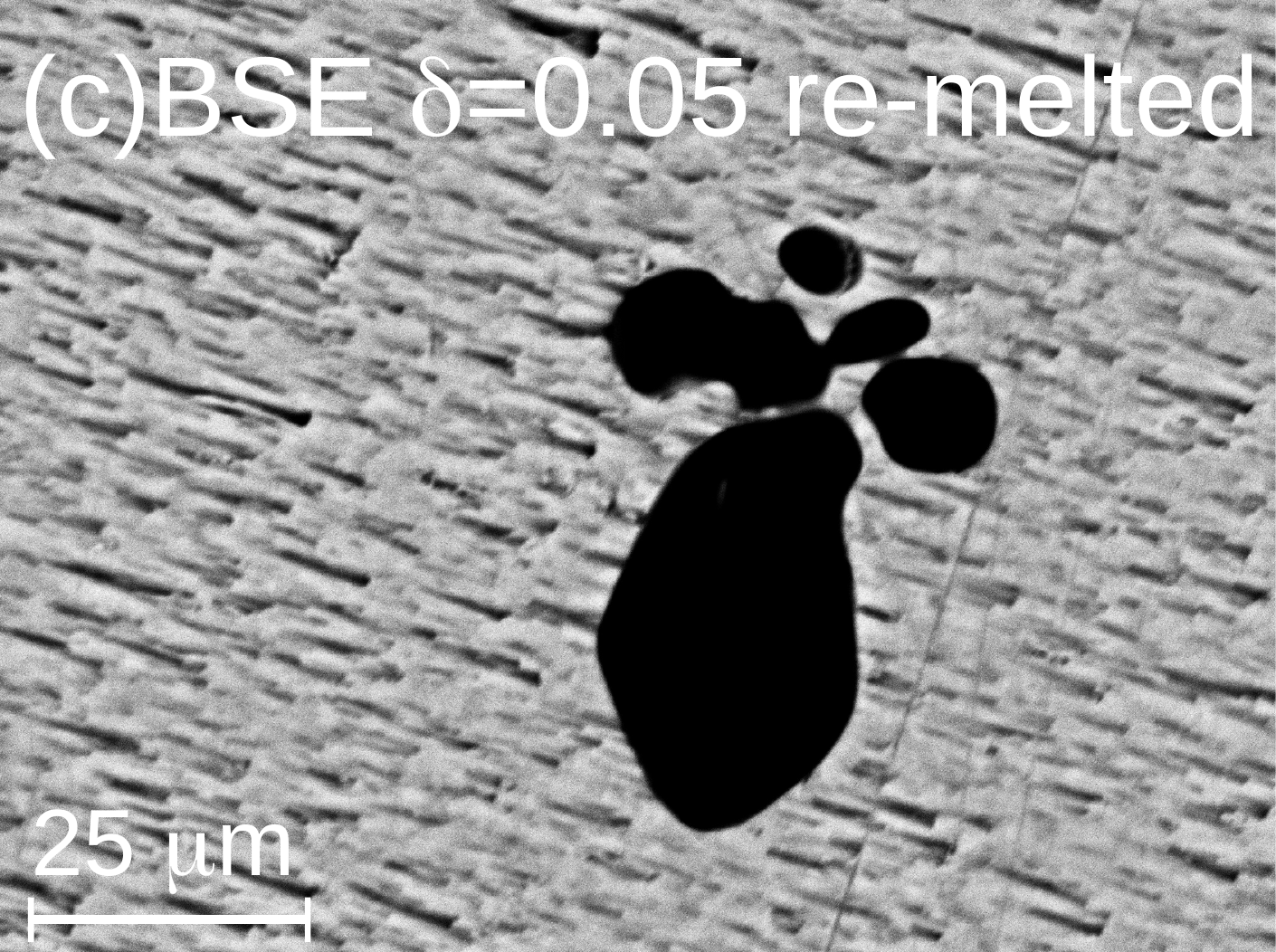}
\includegraphics[width=0.22\textwidth]{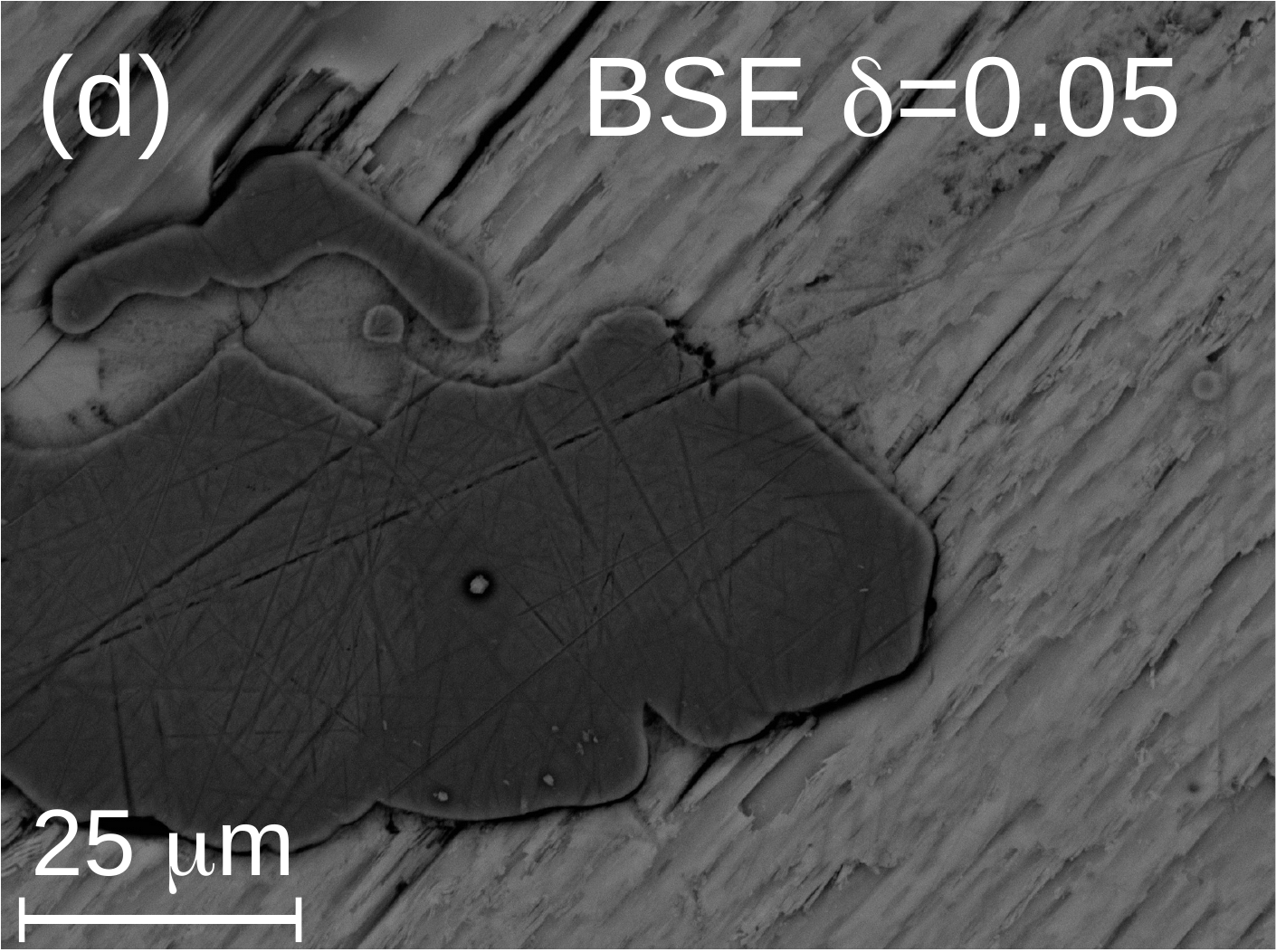}
\includegraphics[width=0.22\textwidth]{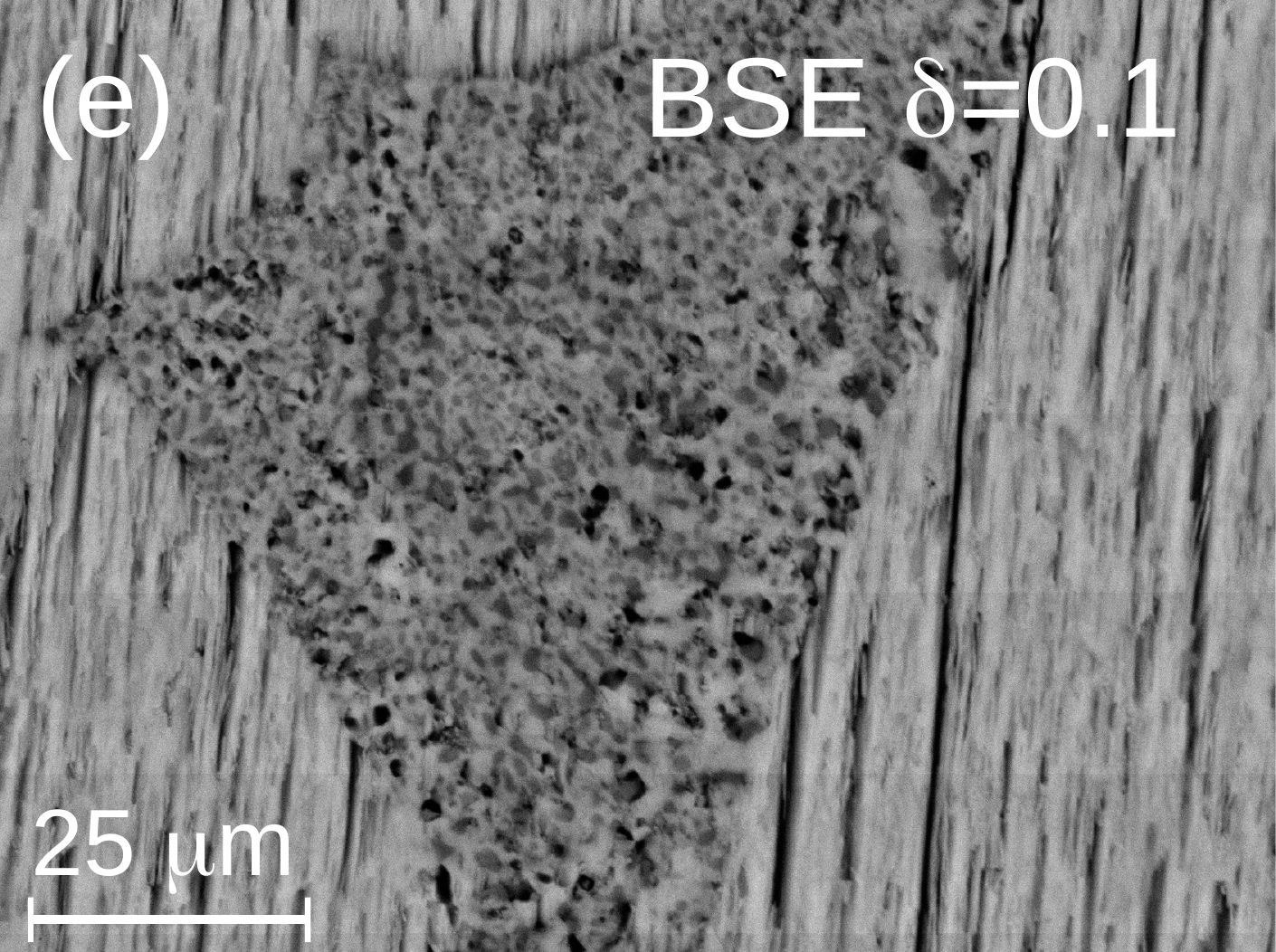}
\includegraphics[width=0.22\textwidth]{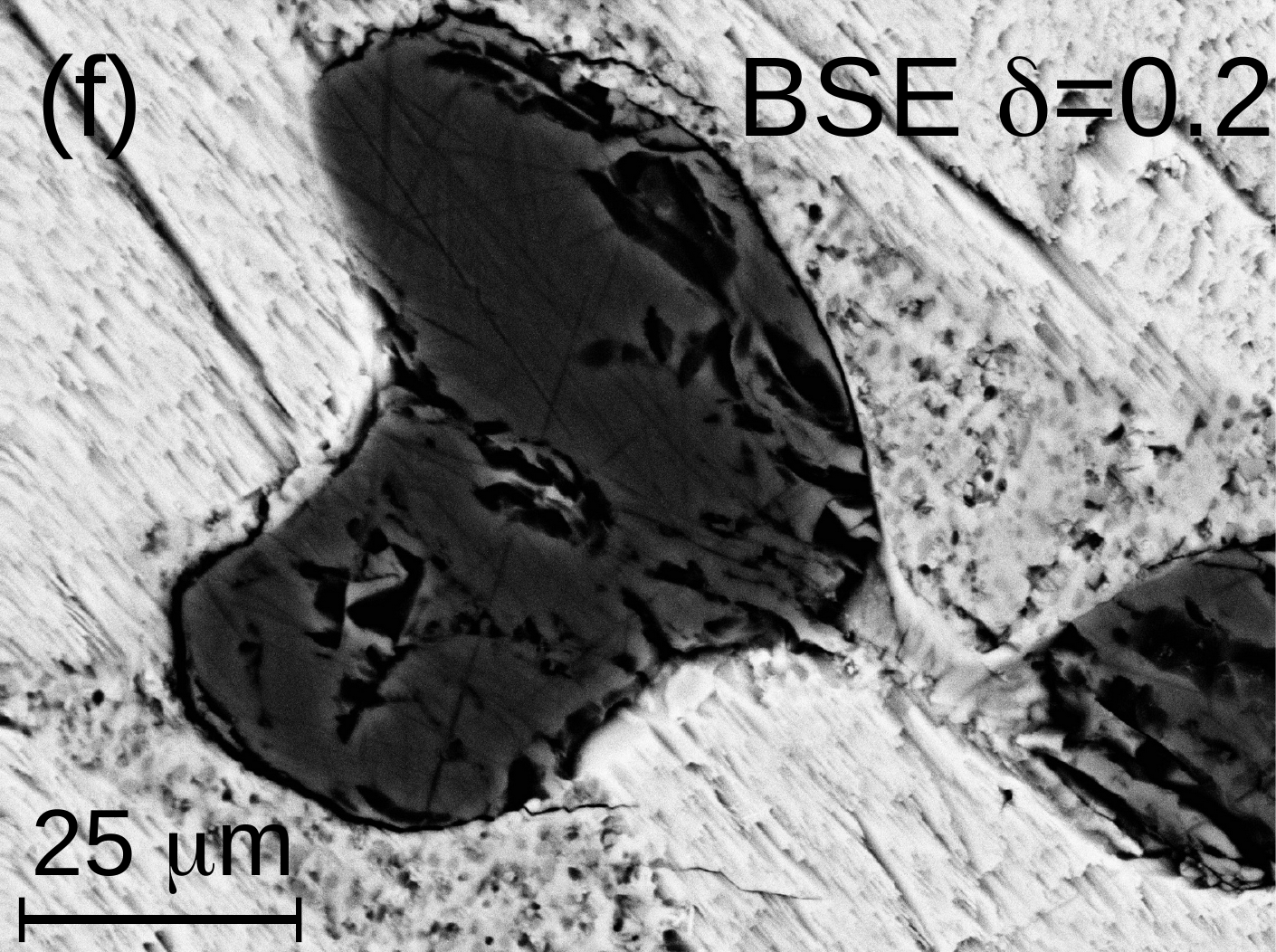}
\end{center}
\caption{(a) Secondary Electron SEM image of BaFe$_{2+\delta}$S$_3$ for $\delta$=0. Back-Scattered Electron SEM images for (b) $\delta$=0 and for inclusions of (c) a Si-rich phase for $\delta$=0.05 (re-melted), (d) Fe for $\delta$=0.05, (e) other ternary phases containing Ba, Fe and S for $\delta$=0.1, and (f) FeS for $\delta$=0.2.}\label{inclusions}
\end{figure}

Secondary Electron (SE) and Back-Scattered Electron (BSE) images for BaFe$_{2+\delta}$S$_3$ samples with nominal $\delta$=0 are shown in Fig. \ref{inclusions}(a) and (b). 
Owing to their quasi-one dimensional structure, the samples are difficult to polish, as attested by the characteristic needle-like striations.  
Composition measured over the large main phase areas within a sample differs by less than 2\%At and shows no monotonic change as a function of the nominal Fe content, the resulting average values are summarized in Table \ref{comp-delta}. The average composition is consistent with BaFe$_2$S$_3$ within the accuracy of the method ($\sim$2\%At). 
It is not clear whether the small dispersion observed within each batch is indicative of chemical inhomogeneity or to small amounts of secondary phases or voids beneath the surface.

\begin{table}[b]
\caption{Average composition of the main phase as a function of the excess Fe in the nominal composition, $\delta$, for BaFe$_{2+\delta}$S$_3$. The numbers in parenthesis correspond to the standard deviation. Expected (theoretical) values are also indicated for $\delta$=0 and 0.2.}
\centering
\begin{tabular}{c c c c c c c}
\hline\hline 
  $\delta$ & {Ba}	& {Fe}	& {S}  \\
\hline
0 (theoretical)	&16.67	&33.33	&50 \\
0	&17.2(2)	&32.9(2)	&49.9(3)  \\
0.05 (re-melted)	&16.9(2)	&33.2(5)	&49.9(4)\\
0.05	&17.1(2)	&32.8(2)	&50.1(4) \\
0.1	&16.8(2)	&32.3(4)	&50.9(5)\\
0.2	&17.0(3)	&32.9(4)	&50.1(6)\\
0.2 (theoretical)	&16.13	&35.48	&48.39 \\
\hline\hline 
\end{tabular}	
\label{comp-delta}
\end{table}

Figures \ref{inclusions}(c)-(f) show typical inclusions found in the samples. FeS was observed in the crystals with a nominal Fe content of $\delta$=0.1 and 0.2. Moreover, Fe was identified as an inclusion for crystals with $\delta$=0.05 and 0.2. Other small-sized areas with varying amounts of Ba, Fe and S were also observed. Some measurements presented a small quantity of O, which could have originated from residual oxygen present in the ampule during growth, or from oxidation of the inclusions during the preparation of the samples for the measurements. No presence of O was observed in the main phase. In the case of $\delta$=0.05 (re-melted) a small quantity of a Si-rich phase was found, which possibly indicates a reaction between the melt and the quartz crucible.

Our chemical analysis suggests that the Fe content in the main phase is practically unchanged and the excess Fe used during synthesis mostly leads to the formation of Fe rich secondary phases.
The batch with $\delta$=0.1 yielded a slightly (less than 1\%At) lower Fe content than the others, but as will be shown below, the physical properties for this batch are not significantly different.

\subsection{Structural characterization}
\label{section:rx}

\subsubsection{Powder x-ray diffraction}

\begin{figure}[t!]
\begin{center}
\includegraphics[width=0.48\textwidth]{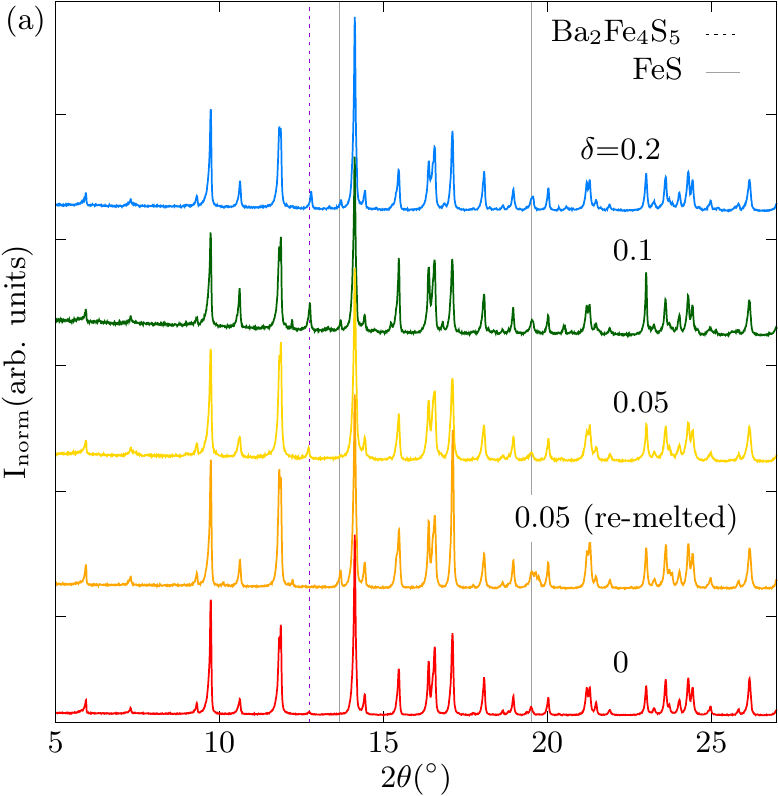}
\includegraphics[width=0.48\textwidth]{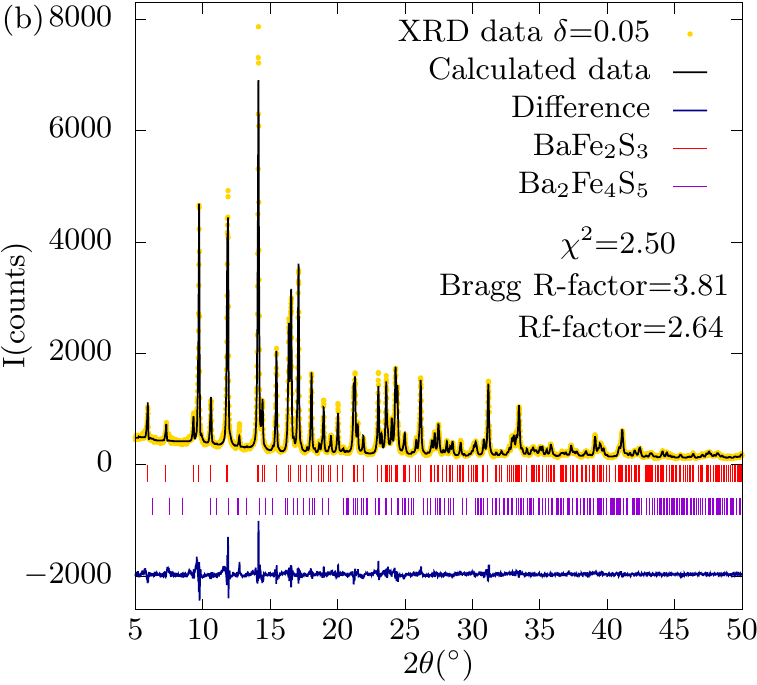}
\end{center}
\caption{(a) Powder x-ray diffraction pattern of BaFe$_{2+\delta}$S$_3$ for nominal $\delta$=0, 0.05, 0.1 and 0.2. Vertical lines indicate the positions of the most intense reflections for the secondary phases. The patterns are vertically shifted for clarity and the intensity is normalized. (b) Rietveld analysis of the powder x-ray diffraction pattern for $\delta$=0.05.}\label{rx-mo}
\end{figure}
Powder x-ray diffraction data for samples with different nominal iron compositions are shown in Fig. \ref{rx-mo}(a). Most reflections can be identified with the orthorhombic structure with space group \textit{Cmcm} (No. 63) previously reported for BaFe$_2$S$_3$ \cite{hong1972crystal}. 
The differences in the relative intensities of the reflections, in particular for $\delta$=0.05 (re-melted) and 0.1, are most probably related to a preferred orientation of the powder, given the highly anisotropic nature of the crystal structure and the needle-like morphology of the crystals.
The lattice parameters were obtained from a Rietveld analysis and are listed in Table \ref{comp-delta-xr}. As an example of the refinement, the case of $\delta$=0.05 is presented in Fig. \ref{rx-mo}(b).
The lattice parameters vary within 0.06-0.1\%, but no systematic change can be observed as a function of $\delta$, as presented in Table \ref{comp-delta-xr}.  
 This is also consistent with the EDX results which show a similar Fe stoichiometry of the main phase for all samples and no correlation with the nominal $\delta$.

\begin{figure}[t!]
\begin{center}
\includegraphics[width=0.48\textwidth]{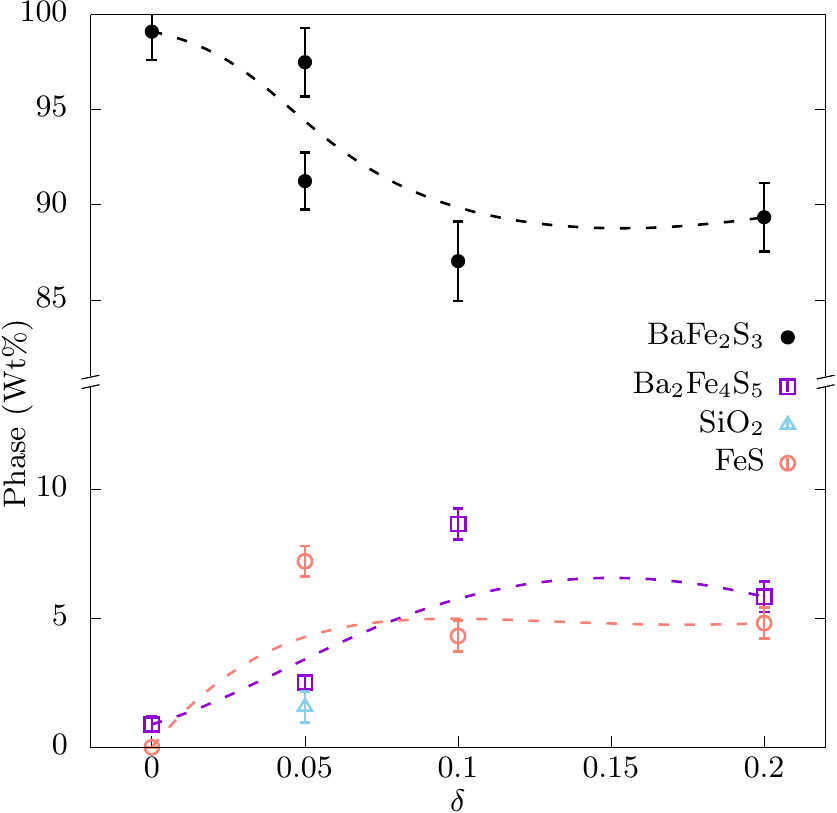}
\end{center}
\caption{Weight fraction of the phases BaFe$_2$S$_3$, Ba$_2$Fe$_4$S$_5$, FeS and SiO$_2$ as a function of the nominal Fe composition. The dotted lines are guides to the eye. }\label{rx-phases}
\end{figure}

Diffraction data confirms the presence of small traces ($<$13\,Wt\%) of impurity phases.
The extra reflections are consistent with Ba$_2$Fe$_4$S$_5$ \cite{swinnea1982crystal} and FeS \cite{marshall2000high}. The most intense reflections of these phases are indicated with lines in Fig. \ref{rx-mo}(a). 
We also identify SiO$_2$ only for the case of $\delta$=0.05 (re-melted). The presence and nature of these inclusions are in good agreement with the EDX measurements. 
From the Rietveld analysis, we obtained that the fraction of overall secondary phase increases with $\delta$, as shown in Fig. \ref{rx-phases}. 
The presence of a small amount of secondary phase also for $\delta$=0 suggests incongruent melting. 

\begin{table}[t!]
\caption{Lattice parameters as a function of the excess Fe in the nominal composition, $\delta$, for BaFe$_{2+\delta}$S$_3$. The number in parenthesis correspond to the error of the fit.}
\centering
\begin{tabular}{c c c c}
\hline\hline 
  $\delta$ & a(\AA) &b(\AA) &c(\AA) \\
\hline
0	&8.7759(2)     &11.2177(3)     &  5.2823(1)   \\
0.05 (re-melted)	&8.7742(4)&11.2137(5)& 5.2793(2)\\
0.05	&8.7762(3)&11.2151(4)&5.2849(2)\\
0.1	&8.7797(4)&11.2211(5)&5.2850(2)\\
0.2	&8.7765(3)& 11.2199(4)&5.2841(2)\\
\hline\hline 
\end{tabular}	
\label{comp-delta-xr}
\end{table} 

\subsubsection{Single crystal x-ray diffraction}

Single crystal x-ray diffraction measurements were conducted to address the question of possible structural and stoichiometry variations of the main phase as a function of $\delta$. For this purpose, single crystals with an edge length of 50\,$\mu$m to 100\,$\mu$m were selected. Table \ref{table:DataCollection} in Appendix \ref{scxr} summarizes parameters of the data collection and the results of the structural refinement. The atomic positions, isotropic and anisotropic displacement parameters are listed in Tables \ref{table:Coordinate} and \ref{table:Displacement}, shown in Appendix \ref{scxr}. No clear monotonic variation of atomic positions, isotropic nor anisotropic displacement parameters is observed.

In line with former studies \cite{hong1972crystal, hirata2015effects} and with the results of the powder x-ray diffraction, the analysis of systematic extinctions and the subsequent structure refinements confirm that BaFe$_2$S$_3$ crystallizes in the orthorhombic space group \textit{Cmcm} (No. 63). Since BaFe$_2$S$_3$ was suggested to be prone to Fe-deficiency \cite{hirata2015effects}, we refined the Fe-occupancy of our single crystals. No significant deviation from the ideal stoichiometry could be detected for our samples; i.e. irrespective of the nominal $\delta$ value, we found BaFe$_{2+\delta}$S$_3$ with $\delta$=0. Furthermore, keeping the site occupancy variable for the remaining atoms did not show a statistically relevant off-stoichiometry. Similar to the closely related Se compound, BaFe$_2$Se$_3$, the maximum residual electron density was observed in the vicinity of the Ba atom. This residual electron density was interpreted earlier as a result of the motion of the weakly bonded Ba atom \cite{Saparov}. This is perfectly in-line with the EDX results described above.
\begin{figure}[b]
\begin{center}
\includegraphics[width=0.48\textwidth]{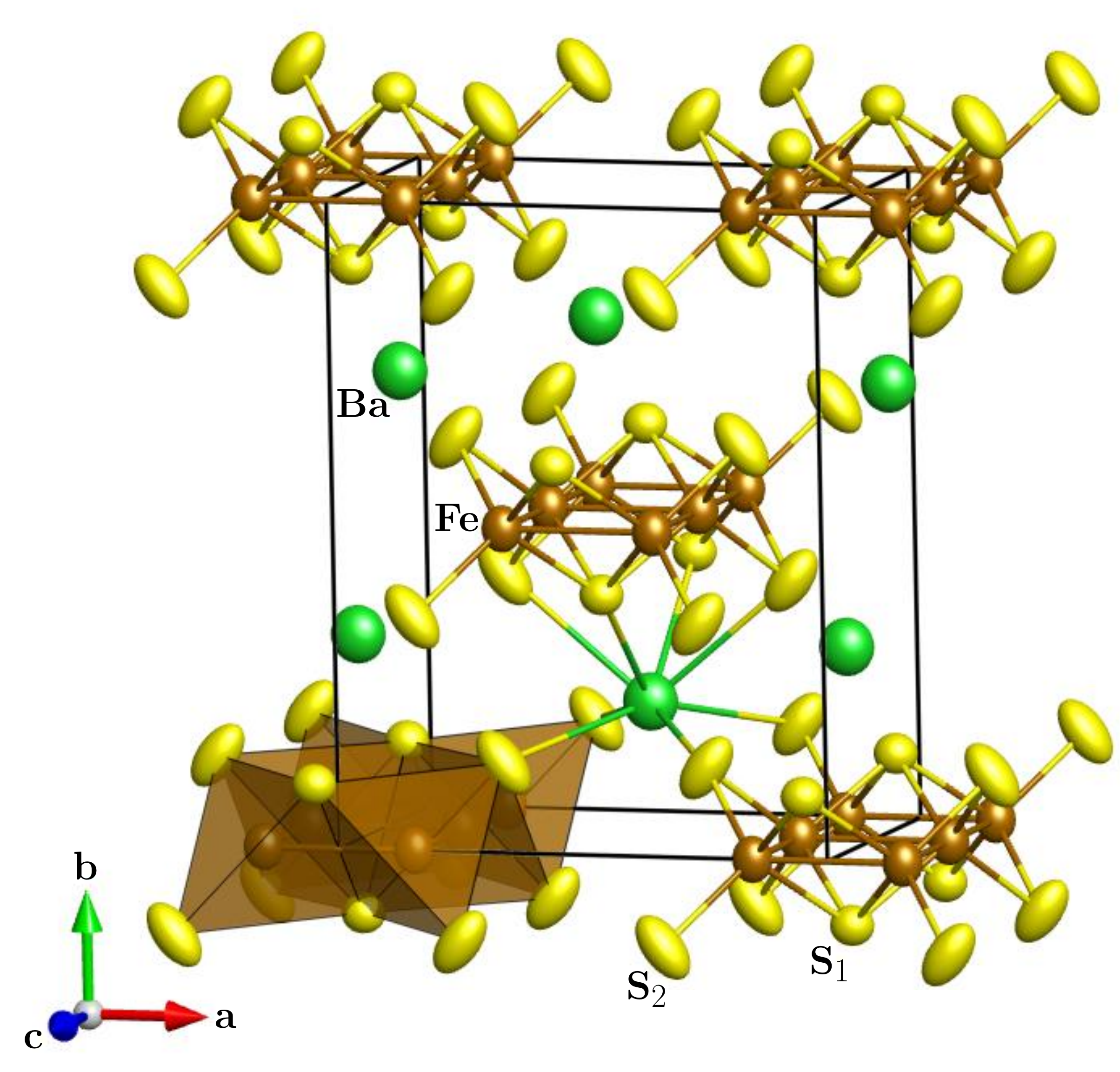}
\end{center}
\caption{Crystal structure of orthorhombic BaFe$_{2}$S$_3$ obtained for $\delta$=0 from single crystal x-ray diffraction measurements. Samples with other $\delta$ values exhibit a very similar structure, see Appendix \ref{scxr}.}
\label{Structure}
\end{figure} 

Figure \ref{Structure} shows the crystal structure of BaFe$_2$S$_3$ based on our refinement. The $c$-axis was found to be parallel to the long direction of the needle-like crystals. The main structural units are two-leg Fe-ladders, assembled by edge-sharing FeS$_4$ tetrahedra, running along the crystallographic \textit{c} direction and channels occupied by Ba atoms. The two Fe-Fe distances in the ladder are $d_{rung}$=2.6976(9)\,$\text{\AA}$  (parallel to \textit{a}) and $d_{leg}$=2.64115(5)\,$\text{\AA}$ (parallel to \textit{c}) for $\delta$=0. These values scatter within 0.1\% as a function of $\delta$. The strong anisotropy of the atomic displacement parameters (ADP) for the S$_{2}$-site, see Appendix \ref{scxr}, is in agreement with the findings by Hong and Steinfink \cite{hong1972crystal}. The long axis of the elongated ``cigar-shaped" ADP ellipsoid lies in the \textit{ab}-plane and is almost perpendicular to the Fe-S$_2$ bond. This may indicate the tendency to break the symmetry by a rotation of the FeS$_4$ tetrahedra within the \textit{ab}-plane similar to the temperature-dependent \textit{Cmcm} to  \textit{Pnma} phase transition in BaFe$_2$Se$_3$ \cite{Svitlyk}.

\subsection{Magnetization}
\label{magnetization}

\begin{figure}[t!]
\begin{center}
\includegraphics[width=0.48\textwidth]{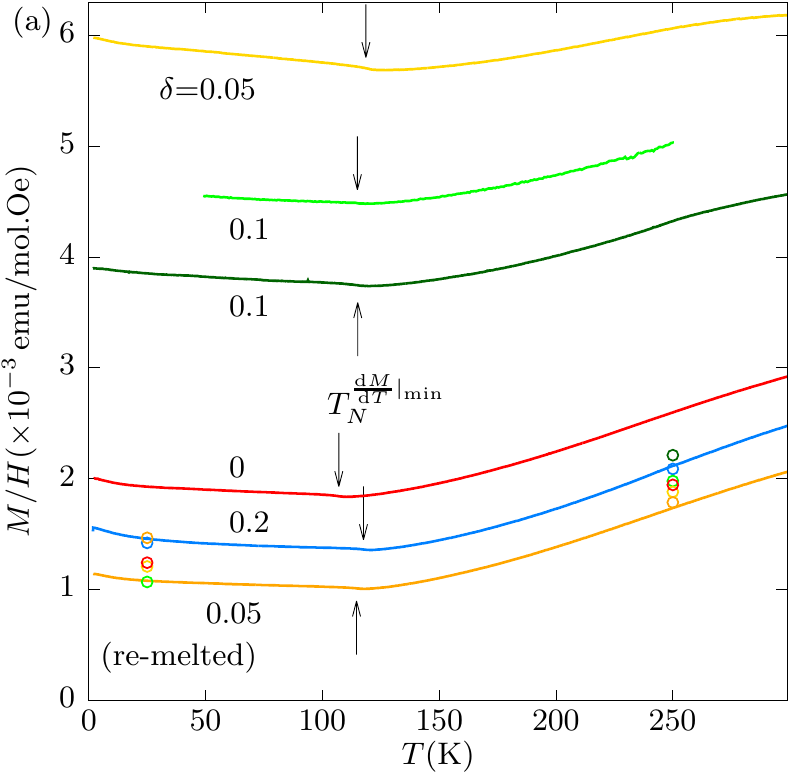}
\includegraphics[width=0.48\textwidth]{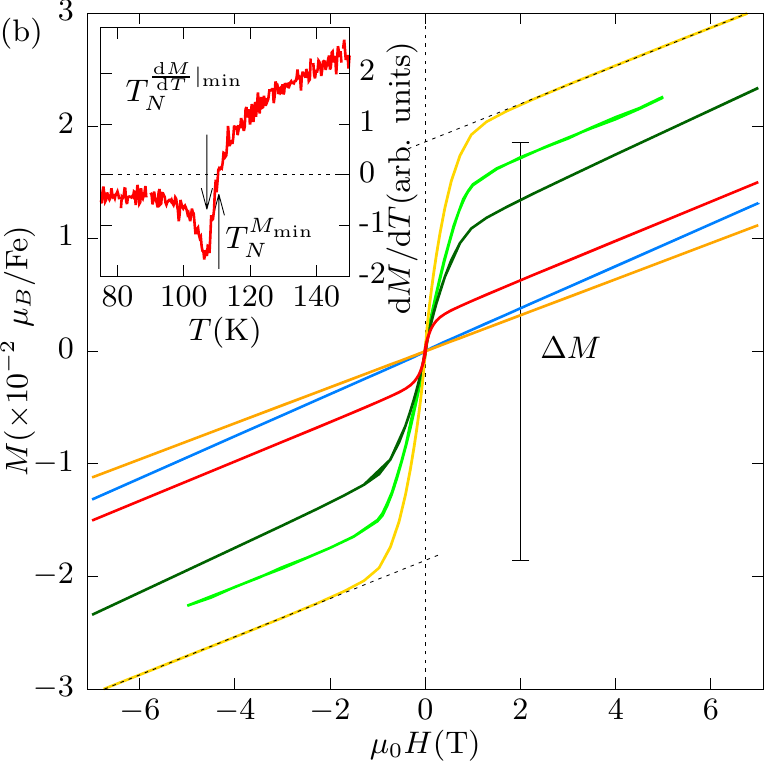}
\end{center}
\caption{(a) Magnetization of BaFe$_{2+\delta}$S$_3$ as a function of temperature for crystals with different nominal Fe content. The arrows indicate the N\'eel temperature as defined in the inset. The circles are the susceptibility obtained from $M$($H$) for fields above 4\,T. The color indicates the value of $\delta$. (b) Field dependence of the magnetization for the same samples of (a) for $T$=250\,K. $\Delta M$ indicates the contribution of extrinsic magnetic inclusions for $\delta$=0.05. Inset: N\'eel temperature definitions for the case of $\delta$=0.}\label{mvst}
\end{figure}

Figure \ref{mvst}(a) presents the magnetization divided by the applied magnetic field, $M/H$, as a function of the temperature, $T$, for different nominal Fe compositions and a magnetic field of $\mu_0H$=5\,T parallel to the $c$-axis. 
We used such a high magnetic field in order to try to saturate any possible ferromagnetic spurious contribution (see below).
Below room temperature, the magnetization decreases with decreasing temperature. 
This tendency, that contrasts the typical Curie-Weiss behavior observed in 3D localized magnets, is independent of $\delta$ and is also commonly observed in other quasi-1D materials \cite{johnston1996antiferromagnetic, tiwary1997single}. 
This behavior can be qualitatively understood if we consider that low-dimensional magnets show short range correlations at temperatures above $T_N$, which will be reflected in a maximum in the susceptibility at a temperature of the order of these correlations. For temperatures below the maximum, one observes a decreasing susceptibility for decreasing temperature.
The change of behavior at around $\sim$120\,K marks the antiferromagnetic transition. 
Below $T_N$, the direction of the applied magnetic field is perpendicular to the magnetic moments as presented in Ref. \cite{takahashi2015pressure}.

It is noticeable that $\delta$ affects the absolute value of $M/H$($T$) in a non-monotonous way, as can be observed in Fig. \ref{mvst}(a). 
This is also reflected in the jump at low fields ($\Delta M$) in the isothermal magnetization for $T$=250\,K shown in Fig. \ref{mvst}(b). 
A similar jump is also found in measurements below $T_N$ (not shown). 
 We find that the size of $\Delta M$ depends in an arbitrary manner on the particular sample, as shown for $\delta$=0.1 in Fig. \ref{mvst}(a) and (b). All this indicates that $\Delta M$ is most probably related to the amount of extrinsic ferromagnetic inclusions present in different amounts in different samples even of the same batch. This kind of behavior was already reported in the literature \cite{zhang2018situ} and was also associated with an extrinsic ferromagnetic contribution.
If we assume that the magnetic inclusion is Fe, it represents less than 2\%mol for all samples. Such a small fraction of Fe most likely remains undetected in powder x-ray diffraction measurements, but is occasionally seen in EDX analysis, see Fig. \ref{inclusions}(d).
It is worth to mention that re-melted samples ($\delta$=0.05) have $\Delta M$$\simeq$0, which indicates that practically no spurious ferromagnetic phase is present. 

In order to get rid of the extrinsic ferromagnetic contribution, the magnetic susceptibility can be calculated from the slope of $M$($H$) well above the characteristic field of the magnetization jump (Honda-Owen analysis, see Ref. \cite{honda}), where the spurious contribution is essentially saturated.  
While $M/H$ varies by a factor of six between samples, the susceptibility (d$M$/d$H$) varies by less than 50\%, see Fig. \ref{mvst}(a). 
The obtained d$M$/d$H$ values (1.07-1.47$\times 10^{-3}$\,emu/mol.Oe for 25\,K) are close to $M/H$($T$) of the re-melted sample.
Therefore, re-melted samples may assumed to be the behavior closest to the intrinsic one of BaFe$_2$S$_3$.

\begin{figure}[t]
\begin{center}
\includegraphics[width=0.48\textwidth]{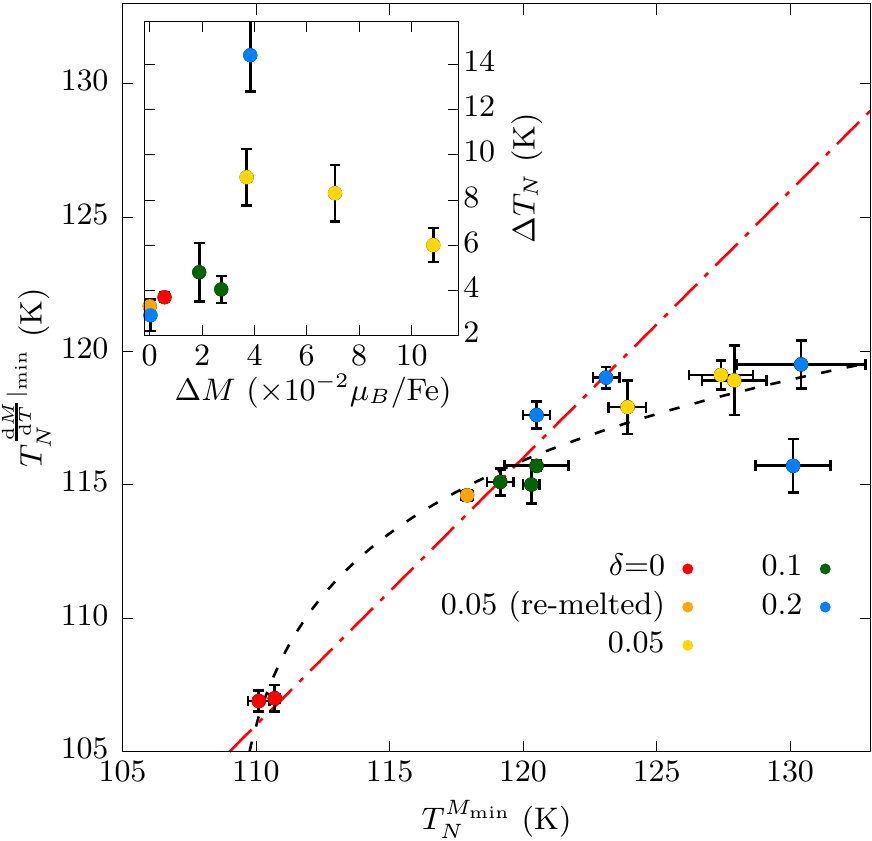}
\end{center}
\caption{Correlation between two definitions of the N\'eel temperature, $T_N^{M_\text{min}}$ and $T_N^{\frac{\text{d}M}{\text{d}T}|_\text{min}}$. The dotted lines are guides to the eye to indicate a linear and non-linear dependence. Inset: $T_N^{M_\text{min}} - T_N^{\frac{\text{d}M}{\text{d}T}|_\text{min}}$ as a function of $\Delta M$ for $T$=250\,K.}\label{correlation}
\end{figure}

However, the value of the susceptibility obtained from the high-field slope of $M$($H$) still presents a small sample dependence that it is neither correlated with the presence of ferromagnetic impurities nor to $\delta$. Further, this variations are too large to be originated in crystal misalignment, especially in the paramagnetic phase \cite{takahashi2015pressure}. Since we know from EDX and powder x-ray diffraction that other magnetic impurity phases may be present (such as antiferromagnetic FeS \cite{hirahara1958magnetic}), these variations in $\text{d}M/\text{d}H$ do not necessarily correspond to variations in the main phase. Further insight can be gained by analyzing the behavior of the N\'eel temperature.

We explored two alternatives for the definition of $T_N$. The first one ($T_N^{M_\text{min}}$) consists in defining $T_N$ as the minimum in the magnetization as a function of the temperature, namely, $\frac{\text{d}M}{\text{d}T}$=0. The second one ($T_N^{\frac{\text{d}M}{\text{d}T}|_\text{min}}$) considers the minimum of the temperature derivative of the magnetization ($\frac{\text{d}M}{\text{d}T}|_\text{min}$), i.e. the maximum negative slope in the curve of $M$($T$) (see inset of Fig.\ref{mvst}(b)). 
Although the two definitions do not result in the same numerical value, they are expected to be closely correlated if they are meaningful indicators of the magnetic phase transition of the main phase. 
Figure \ref{correlation} shows that this is not the case. Both definitions are not equivalent, while there is an appreciable dispersion in $T_N^{M_\text{min}}$ for different samples, even of the same batch, $T_N^{\frac{\text{d}M}{\text{d}T}|_\text{min}}$ is less sensitive to sample variations within a batch. 
The samples with the least amount of ferromagnetic impurity exhibit the smallest difference between both definitions of $T_N$ and this difference is largest for samples with higher values of $T_N^{M_\text{min}}$, see inset of Fig. \ref{correlation}. This suggests that the value of $T_N^{M_\text{min}}$ is more susceptible to the ferromagnetic impurities than $T_N^{\frac{\text{d}M}{\text{d}T}|_\text{min}}$ and that a high value of $T_N^{M_\text{min}}$ does not necessarily correspond to a higher chemical purity of the sample, in contrast to what is commonly assumed.
Irrespective of the definition chosen, there is no clear dependence of the value of $T_N$ and $\delta$, with the exception of the samples with $\delta$=0 that present the lowest value of $T_N$. However, no difference in the main phase was previously observed for $\delta$=0 in the structural and compositional analysis.

\subsection{Resistivity}
\label{resistivity}

\begin{figure}[h]
\begin{center}
\includegraphics[width=0.48\textwidth]{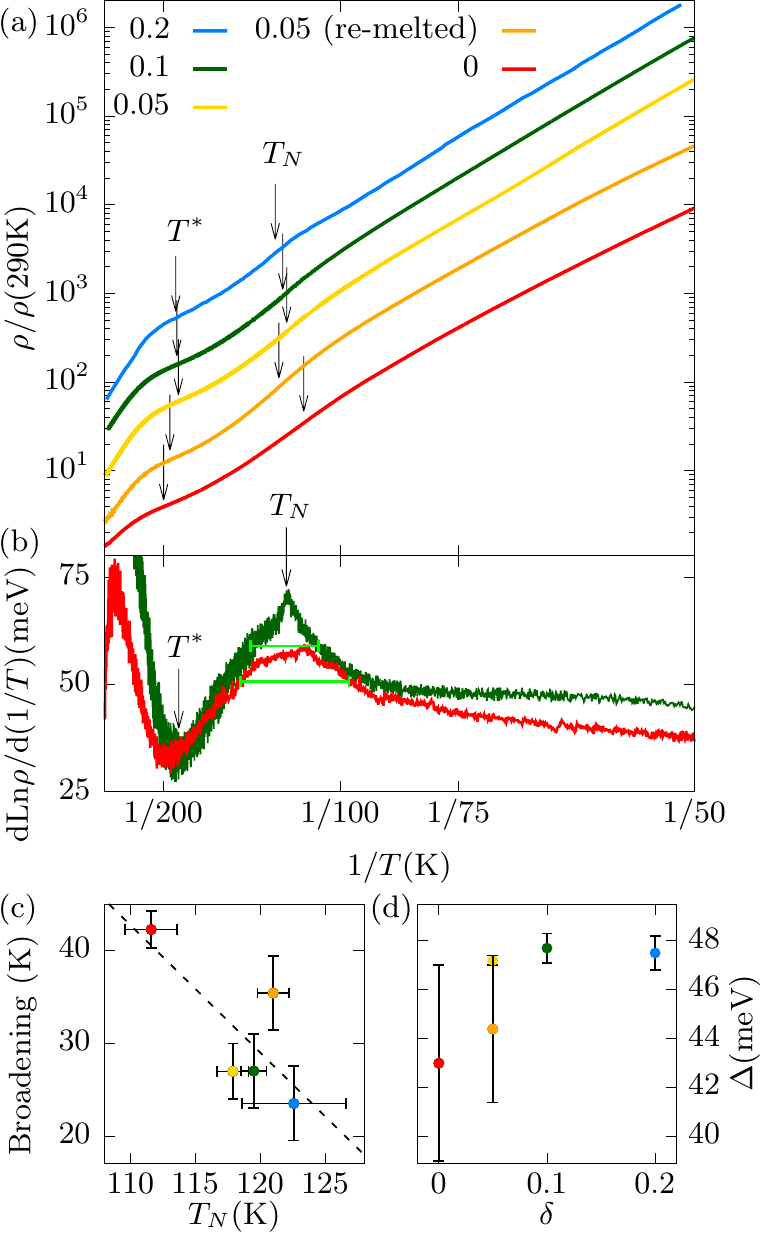}
\end{center}
\caption{(a) Normalized resistivity as a function of the inverse temperature for different nominal values of $\delta$ for BaFe$_{2+\delta}$S$_3$.  The curves are vertically shifted for clarity. (b) Temperature dependence of the derivative of the logarithmic of the resistivity with respect to the inverse temperature for $\delta$=0.1 and 0. $T_N$ and $T^\ast$ are indicated with an arrow for $\delta$=0.1. The horizontal segments indicate the broadening of the curves around $T_N$, computed as the width of the curves at half of the distance from their maximum to the value of $\Delta$. (c) Broadening as a function of $T_N$. The dotted line is a guide to the eye. (d) Value of the energy gap for $T<T_N$ as a function of the nominal Fe content.}\label{rvst}
\end{figure}

Figure \ref{rvst}(a) presents the resistivity as a function of the inverse temperature for samples with different nominal Fe composition. The resistivity increases with decreasing temperature indicating an insulating behavior. This behavior is shared by all samples in spite of the different nominal $\delta$ value. Two characteristic temperatures can be identified, the N\'eel temperature at $T_N$$\sim$120\,K and a slope change at $T^*$$\sim$190\,K, as indicated in Fig. \ref{rvst}(a) for different nominal values of $\delta$. The definition used for these characteristic temperatures is shown in Fig. \ref{rvst}(b). The origin of the change at $T^*$ is still under debate, but it has been suggested to be related to an orbital ordering transition \cite{yamauchi2015pressure, hirata2015effects}. 
The samples with $\delta$=0 present the lowest value of $T_N$, as already observed in magnetization measurements, and a visibly larger broadening in the area around $T_N$ as shown in Fig. \ref{rvst}(b) and \ref{rvst}(c).

\begin{figure}[t]
\begin{center}
\includegraphics[width=0.48\textwidth]{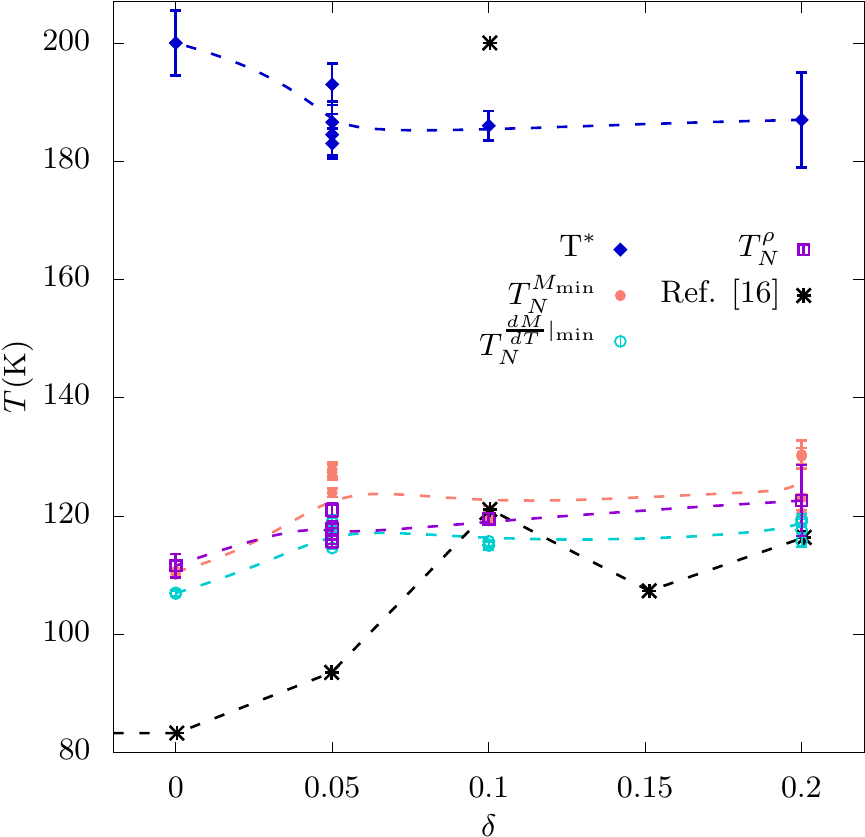}
\end{center}
\caption{Dependence of $T_N$ and $T^*$ with the nominal Fe content for BaFe$_{2+\delta}$S$_3$. $T_N^{M_\text{min}}$, $T_N^{\frac{\text{d}M}{\text{d}T}|_\text{min}}$ and $T_N^\rho$ correspond to the N\'eel temperature obtained from magnetization (inset of Fig. \ref{mvst}(b)) and resistivity measurements (Fig. \ref{rvst}), respectively. For comparison, $T_N$ and $T^*$ extracted from reference \cite{hirata2015effects} are also included. The dotted lines are guides to the eye.}\label{diagram}
\end{figure}

For $\delta$$\geq$0.05 and $T$$\lesssim$90\,K, dLn$\rho$/d$(1/T)$ is practically temperature independent, which allows us to describe the resistivity with a thermally activated behavior ($\rho=\rho_0e^{\Delta/T}$), as shown in Fig. \ref{rvst}(b) for $\delta$=0.1 as a representative curve. The same kind of behavior is also observed by Hirata \textit{et al.} \cite{hirata2015effects} but only for $\delta$=0.1.
The resulting gap ($\Delta$) from our data for different $\delta$ values is plotted in Fig. \ref{rvst}(d), and is in a good agreement with the value for $\delta$=0.1 from the literature (47\,meV) \cite{hirata2015effects}.
For $\delta$=0 and 0.05 (re-melted), the resistivity slightly deviates from a thermally activated behavior. 
For completeness, the obtained gap in these cases using as fitting interval the range 60\,K to 90\,K is also included in Fig. \ref{rvst}(d).
A 1D variable range hopping, $\rho=\rho_0e^{(T_0/T)^{1/2}}$ \cite{fogler2004variable}, is found to quantitatively fit the resistivity below $\sim$90\,K for $\delta$=0.
This kind of behavior has been previously observed in the literature \cite{zhang2018situ, hirata2015effects}. 

Figure \ref{diagram} presents the dependence of the N\'eel temperature and $T^*$ with the nominal Fe content. $T_N$ as estimated from resistivity measurements (as defined in Fig. \ref{rvst}) and from the magnetization (inset Fig. \ref{mvst}(b)) presents only a weak increase with $\delta$.
This contrasts with Ref. \cite{hirata2015effects}, where a change in $T_N$ of $\sim$40\,K was reported with a maximum for $\delta$=0.1. 
In addition, $T^*$ is present for all samples and is seemingly $\delta$-independent for $\delta \geq 0.05$.
The crystals with $\delta$=0 present the lowest value of $T_N$ and the highest value of $T^*$. This fact, together with the larger broadening in the resistivity curves around $T_N$ (Fig. \ref{rvst}(c)) suggests the presence of structural defects other than Fe occupancy. However, no sign of structural defects is seen in the compositional and structural characterization.

\section{Conclusions}
\label{conclusions}

We have studied the effect of the nominal Fe excess in BaFe$_{2+\delta}$S$_3$ in the composition, the crystal structure, the magnetization and the resistivity. The analysis of the composition, as well as the crystal structure, obtained from powder and single crystal x-ray diffraction measurements, indicate that the extra Fe induces the formation of Fe-rich inclusions rather than being incorporated in the structure. The nature of these extra phases depends on the value of $\delta$ and they are not homogeneously distributed within a batch. 
Magnetic inclusions affect the experimentally observed magnetization curves, giving rise to higher absolute values of $M/H$ and a magnetization step in $M$($H$).
Importantly, we found that such ferromagnetic inclusions can be excluded by directional crystallization of remelted samples using a Bridgman procedure.
The magnetization and the resistivity are consistent with an antiferromagnetic transition at $T_N\sim$120\,K. The value of $T_N$ weakly depends on $\delta$. 
Interestingly, samples with nominal $\delta$=0 present the smallest $T_N$ and the highest $T^*$. 
For $\delta$$\geq$0.05, the analysis of the transport and magnetic measurements is consistent with the composition and structural characterization in that the main role of the excess Fe in the growth procedure is forming impurity phases and not modifying the intrinsic behavior of BaFe$_{2}$S$_3$.

\section{Acknowledgments}
We thank G. Kreutzer and S. M\"uller-Litvanyi for the assistance in the EDX measurements, C.G.F. Blum for the assistance in the powder x-ray measurements, S. Gass and G. Bastien for technical assistance in the magnetization measurements and F. Caglieris and C. Wuttke for technical assistance in the resistivity measurements.
This research has been supported by the Deutsche Forschungsgemeinschaft (DFG) through SFB 1143 (Project No. 247310070), the DFG within the Graduate School GRK 1621 and the W\"urzburg-Dresden Cluster of Excellence on Complexity and Topology in Quantum Matter – ct.qmat (EXC 2147, Project No. 390858490).
M.L.A. acknowledges support from the Alexander von Humboldt Foundation through the Georg Forster program. 

\bibliography{mibib}{}

\begin{thebibliography}{34}%
\makeatletter
\providecommand \@ifxundefined [1]{%
 \@ifx{#1\undefined}
}%
\providecommand \@ifnum [1]{%
 \ifnum #1\expandafter \@firstoftwo
 \else \expandafter \@secondoftwo
 \fi
}%
\providecommand \@ifx [1]{%
 \ifx #1\expandafter \@firstoftwo
 \else \expandafter \@secondoftwo
 \fi
}%
\providecommand \natexlab [1]{#1}%
\providecommand \enquote  [1]{``#1''}%
\providecommand \bibnamefont  [1]{#1}%
\providecommand \bibfnamefont [1]{#1}%
\providecommand \citenamefont [1]{#1}%
\providecommand \href@noop [0]{\@secondoftwo}%
\providecommand \href [0]{\begingroup \@sanitize@url \@href}%
\providecommand \@href[1]{\@@startlink{#1}\@@href}%
\providecommand \@@href[1]{\endgroup#1\@@endlink}%
\providecommand \@sanitize@url [0]{\catcode `\\12\catcode `\$12\catcode
  `\&12\catcode `\#12\catcode `\^12\catcode `\_12\catcode `\%12\relax}%
\providecommand \@@startlink[1]{}%
\providecommand \@@endlink[0]{}%
\providecommand \url  [0]{\begingroup\@sanitize@url \@url }%
\providecommand \@url [1]{\endgroup\@href {#1}{\urlprefix }}%
\providecommand \urlprefix  [0]{URL }%
\providecommand \Eprint [0]{\href }%
\providecommand \doibase [0]{http://dx.doi.org/}%
\providecommand \selectlanguage [0]{\@gobble}%
\providecommand \bibinfo  [0]{\@secondoftwo}%
\providecommand \bibfield  [0]{\@secondoftwo}%
\providecommand \translation [1]{[#1]}%
\providecommand \BibitemOpen [0]{}%
\providecommand \bibitemStop [0]{}%
\providecommand \bibitemNoStop [0]{.\EOS\space}%
\providecommand \EOS [0]{\spacefactor3000\relax}%
\providecommand \BibitemShut  [1]{\csname bibitem#1\endcsname}%
\let\auto@bib@innerbib\@empty
\bibitem [{\citenamefont {Takahashi}\ \emph {et~al.}(2015)\citenamefont
  {Takahashi}, \citenamefont {Sugimoto}, \citenamefont {Nambu}, \citenamefont
  {Yamauchi}, \citenamefont {Hirata}, \citenamefont {Kawakami}, \citenamefont
  {Avdeev}, \citenamefont {Matsubayashi}, \citenamefont {Du}, \citenamefont
  {Kawashima} \emph {et~al.}}]{takahashi2015pressure}%
  \BibitemOpen
  \bibfield  {author} {\bibinfo {author} {\bibfnamefont {H.}~\bibnamefont
  {Takahashi}}, \bibinfo {author} {\bibfnamefont {A.}~\bibnamefont {Sugimoto}},
  \bibinfo {author} {\bibfnamefont {Y.}~\bibnamefont {Nambu}}, \bibinfo
  {author} {\bibfnamefont {T.}~\bibnamefont {Yamauchi}}, \bibinfo {author}
  {\bibfnamefont {Y.}~\bibnamefont {Hirata}}, \bibinfo {author} {\bibfnamefont
  {T.}~\bibnamefont {Kawakami}}, \bibinfo {author} {\bibfnamefont
  {M.}~\bibnamefont {Avdeev}}, \bibinfo {author} {\bibfnamefont
  {K.}~\bibnamefont {Matsubayashi}}, \bibinfo {author} {\bibfnamefont
  {F.}~\bibnamefont {Du}}, \bibinfo {author} {\bibfnamefont {C.}~\bibnamefont
  {Kawashima}},  \emph {et~al.},\ }\href {https://doi.org/10.1038/nmat4351}
  {\bibfield  {journal} {\bibinfo  {journal} {Nature materials}\ }\textbf
  {\bibinfo {volume} {14}},\ \bibinfo {pages} {1008} (\bibinfo {year}
  {2015})}\BibitemShut {NoStop}%
\bibitem [{\citenamefont {Ying}\ \emph {et~al.}(2017)\citenamefont {Ying},
  \citenamefont {Lei}, \citenamefont {Petrovic}, \citenamefont {Xiao},\ and\
  \citenamefont {Struzhkin}}]{ying2017interplay}%
  \BibitemOpen
  \bibfield  {author} {\bibinfo {author} {\bibfnamefont {J.}~\bibnamefont
  {Ying}}, \bibinfo {author} {\bibfnamefont {H.}~\bibnamefont {Lei}}, \bibinfo
  {author} {\bibfnamefont {C.}~\bibnamefont {Petrovic}}, \bibinfo {author}
  {\bibfnamefont {Y.}~\bibnamefont {Xiao}}, \ and\ \bibinfo {author}
  {\bibfnamefont {V.~V.}\ \bibnamefont {Struzhkin}},\ }\href
  {https://link.aps.org/doi/10.1103/PhysRevB.95.241109} {\bibfield  {journal}
  {\bibinfo  {journal} {Physical Review B}\ }\textbf {\bibinfo {volume} {95}},\
  \bibinfo {pages} {241109} (\bibinfo {year} {2017})}\BibitemShut {NoStop}%
\bibitem [{\citenamefont {Yamauchi}\ \emph {et~al.}(2015)\citenamefont
  {Yamauchi}, \citenamefont {Hirata}, \citenamefont {Ueda},\ and\ \citenamefont
  {Ohgushi}}]{yamauchi2015pressure}%
  \BibitemOpen
  \bibfield  {author} {\bibinfo {author} {\bibfnamefont {T.}~\bibnamefont
  {Yamauchi}}, \bibinfo {author} {\bibfnamefont {Y.}~\bibnamefont {Hirata}},
  \bibinfo {author} {\bibfnamefont {Y.}~\bibnamefont {Ueda}}, \ and\ \bibinfo
  {author} {\bibfnamefont {K.}~\bibnamefont {Ohgushi}},\ }\href
  {https://link.aps.org/doi/10.1103/PhysRevLett.115.246402} {\bibfield
  {journal} {\bibinfo  {journal} {Physical Review Letters}\ }\textbf {\bibinfo
  {volume} {115}},\ \bibinfo {pages} {246402} (\bibinfo {year}
  {2015})}\BibitemShut {NoStop}%
\bibitem [{\citenamefont {Hong}\ and\ \citenamefont
  {Steinfink}(1972)}]{hong1972crystal}%
  \BibitemOpen
  \bibfield  {author} {\bibinfo {author} {\bibfnamefont {H.}~\bibnamefont
  {Hong}}\ and\ \bibinfo {author} {\bibfnamefont {H.}~\bibnamefont
  {Steinfink}},\ }\href {https://doi.org/10.1016/0022-4596(72)90015-1}
  {\bibfield  {journal} {\bibinfo  {journal} {Journal of Solid State
  Chemistry}\ }\textbf {\bibinfo {volume} {5}},\ \bibinfo {pages} {93}
  (\bibinfo {year} {1972})}\BibitemShut {NoStop}%
\bibitem [{\citenamefont {Dagotto}\ and\ \citenamefont
  {Rice}(1996)}]{dagotto1996surprises}%
  \BibitemOpen
  \bibfield  {author} {\bibinfo {author} {\bibfnamefont {E.}~\bibnamefont
  {Dagotto}}\ and\ \bibinfo {author} {\bibfnamefont {T.}~\bibnamefont {Rice}},\
  }\href {https://doi.org/10.1126/science.271.5249.618} {\bibfield  {journal}
  {\bibinfo  {journal} {Science}\ }\textbf {\bibinfo {volume} {271}},\ \bibinfo
  {pages} {618} (\bibinfo {year} {1996})}\BibitemShut {NoStop}%
\bibitem [{\citenamefont {Magishi}\ \emph {et~al.}(1998)\citenamefont
  {Magishi}, \citenamefont {Matsumoto}, \citenamefont {Kitaoka}, \citenamefont
  {Ishida}, \citenamefont {Asayama}, \citenamefont {Uehara}, \citenamefont
  {Nagata},\ and\ \citenamefont {Akimitsu}}]{magishi1998spin}%
  \BibitemOpen
  \bibfield  {author} {\bibinfo {author} {\bibfnamefont {K.}~\bibnamefont
  {Magishi}}, \bibinfo {author} {\bibfnamefont {S.}~\bibnamefont {Matsumoto}},
  \bibinfo {author} {\bibfnamefont {Y.}~\bibnamefont {Kitaoka}}, \bibinfo
  {author} {\bibfnamefont {K.}~\bibnamefont {Ishida}}, \bibinfo {author}
  {\bibfnamefont {K.}~\bibnamefont {Asayama}}, \bibinfo {author} {\bibfnamefont
  {M.}~\bibnamefont {Uehara}}, \bibinfo {author} {\bibfnamefont
  {T.}~\bibnamefont {Nagata}}, \ and\ \bibinfo {author} {\bibfnamefont
  {J.}~\bibnamefont {Akimitsu}},\ }\href
  {https://link.aps.org/doi/10.1103/PhysRevB.57.11533} {\bibfield  {journal}
  {\bibinfo  {journal} {Physical Review B}\ }\textbf {\bibinfo {volume} {57}},\
  \bibinfo {pages} {11533} (\bibinfo {year} {1998})}\BibitemShut {NoStop}%
\bibitem [{\citenamefont {Craco}\ and\ \citenamefont
  {Leoni}(2018)}]{craco2018microscopic}%
  \BibitemOpen
  \bibfield  {author} {\bibinfo {author} {\bibfnamefont {L.}~\bibnamefont
  {Craco}}\ and\ \bibinfo {author} {\bibfnamefont {S.}~\bibnamefont {Leoni}},\
  }\href {https://link.aps.org/doi/10.1103/PhysRevB.98.195107} {\bibfield
  {journal} {\bibinfo  {journal} {Physical Review B}\ }\textbf {\bibinfo
  {volume} {98}},\ \bibinfo {pages} {195107} (\bibinfo {year}
  {2018})}\BibitemShut {NoStop}%
\bibitem [{\citenamefont {Roh}\ \emph {et~al.}(2020)\citenamefont {Roh},
  \citenamefont {Shin}, \citenamefont {Jang}, \citenamefont {Lee},
  \citenamefont {Lee}, \citenamefont {Seo}, \citenamefont {Li}, \citenamefont
  {Biesner}, \citenamefont {Dressel}, \citenamefont {Rhee} \emph
  {et~al.}}]{roh2020magnetic}%
  \BibitemOpen
  \bibfield  {author} {\bibinfo {author} {\bibfnamefont {S.}~\bibnamefont
  {Roh}}, \bibinfo {author} {\bibfnamefont {S.}~\bibnamefont {Shin}}, \bibinfo
  {author} {\bibfnamefont {J.}~\bibnamefont {Jang}}, \bibinfo {author}
  {\bibfnamefont {S.}~\bibnamefont {Lee}}, \bibinfo {author} {\bibfnamefont
  {M.}~\bibnamefont {Lee}}, \bibinfo {author} {\bibfnamefont {Y.-S.}\
  \bibnamefont {Seo}}, \bibinfo {author} {\bibfnamefont {W.}~\bibnamefont
  {Li}}, \bibinfo {author} {\bibfnamefont {T.}~\bibnamefont {Biesner}},
  \bibinfo {author} {\bibfnamefont {M.}~\bibnamefont {Dressel}}, \bibinfo
  {author} {\bibfnamefont {J.~Y.}\ \bibnamefont {Rhee}},  \emph {et~al.},\
  }\href {https://link.aps.org/doi/10.1103/PhysRevB.101.115118} {\bibfield
  {journal} {\bibinfo  {journal} {Physical Review B}\ }\textbf {\bibinfo
  {volume} {101}},\ \bibinfo {pages} {115118} (\bibinfo {year}
  {2020})}\BibitemShut {NoStop}%
\bibitem [{\citenamefont {Pizarro}\ and\ \citenamefont
  {Bascones}(2019)}]{PhysRevMaterials3014801}%
  \BibitemOpen
  \bibfield  {author} {\bibinfo {author} {\bibfnamefont {J.~M.}\ \bibnamefont
  {Pizarro}}\ and\ \bibinfo {author} {\bibfnamefont {E.}~\bibnamefont
  {Bascones}},\ }\href {\doibase 10.1103/PhysRevMaterials.3.014801} {\bibfield
  {journal} {\bibinfo  {journal} {Physical Review Materials}\ }\textbf
  {\bibinfo {volume} {3}},\ \bibinfo {pages} {014801} (\bibinfo {year}
  {2019})}\BibitemShut {NoStop}%
\bibitem [{\citenamefont {Zhang}\ \emph {et~al.}(2017)\citenamefont {Zhang},
  \citenamefont {Lin}, \citenamefont {Zhang}, \citenamefont {Dagotto},\ and\
  \citenamefont {Dong}}]{PhysRevB95115154}%
  \BibitemOpen
  \bibfield  {author} {\bibinfo {author} {\bibfnamefont {Y.}~\bibnamefont
  {Zhang}}, \bibinfo {author} {\bibfnamefont {L.}~\bibnamefont {Lin}}, \bibinfo
  {author} {\bibfnamefont {J.-J.}\ \bibnamefont {Zhang}}, \bibinfo {author}
  {\bibfnamefont {E.}~\bibnamefont {Dagotto}}, \ and\ \bibinfo {author}
  {\bibfnamefont {S.}~\bibnamefont {Dong}},\ }\href {\doibase
  10.1103/PhysRevB.95.115154} {\bibfield  {journal} {\bibinfo  {journal}
  {Physical Review B}\ }\textbf {\bibinfo {volume} {95}},\ \bibinfo {pages}
  {115154} (\bibinfo {year} {2017})}\BibitemShut {NoStop}%
\bibitem [{\citenamefont {Materne}\ \emph {et~al.}(2019)\citenamefont
  {Materne}, \citenamefont {Bi}, \citenamefont {Zhao}, \citenamefont {Hu},
  \citenamefont {Amig\'o}, \citenamefont {Seiro}, \citenamefont {Aswartham},
  \citenamefont {B\"uchner},\ and\ \citenamefont
  {Alp}}]{materne2018bandwidth1}%
  \BibitemOpen
  \bibfield  {author} {\bibinfo {author} {\bibfnamefont {P.}~\bibnamefont
  {Materne}}, \bibinfo {author} {\bibfnamefont {W.}~\bibnamefont {Bi}},
  \bibinfo {author} {\bibfnamefont {J.}~\bibnamefont {Zhao}}, \bibinfo {author}
  {\bibfnamefont {M.~Y.}\ \bibnamefont {Hu}}, \bibinfo {author} {\bibfnamefont
  {M.~L.}\ \bibnamefont {Amig\'o}}, \bibinfo {author} {\bibfnamefont
  {S.}~\bibnamefont {Seiro}}, \bibinfo {author} {\bibfnamefont
  {S.}~\bibnamefont {Aswartham}}, \bibinfo {author} {\bibfnamefont
  {B.}~\bibnamefont {B\"uchner}}, \ and\ \bibinfo {author} {\bibfnamefont
  {E.~E.}\ \bibnamefont {Alp}},\ }\href {\doibase 10.1103/PhysRevB.99.020505}
  {\bibfield  {journal} {\bibinfo  {journal} {Physical Review B}\ }\textbf
  {\bibinfo {volume} {99}},\ \bibinfo {pages} {020505} (\bibinfo {year}
  {2019})}\BibitemShut {NoStop}%
\bibitem [{\citenamefont {Zheng}\ \emph {et~al.}(2018)\citenamefont {Zheng},
  \citenamefont {Frandsen}, \citenamefont {Wu}, \citenamefont {Yi},
  \citenamefont {Wu}, \citenamefont {Huang}, \citenamefont
  {Bourret-Courchesne}, \citenamefont {Simutis}, \citenamefont {Khasanov},
  \citenamefont {Yao} \emph {et~al.}}]{zheng2018gradual}%
  \BibitemOpen
  \bibfield  {author} {\bibinfo {author} {\bibfnamefont {L.}~\bibnamefont
  {Zheng}}, \bibinfo {author} {\bibfnamefont {B.~A.}\ \bibnamefont {Frandsen}},
  \bibinfo {author} {\bibfnamefont {C.}~\bibnamefont {Wu}}, \bibinfo {author}
  {\bibfnamefont {M.}~\bibnamefont {Yi}}, \bibinfo {author} {\bibfnamefont
  {S.}~\bibnamefont {Wu}}, \bibinfo {author} {\bibfnamefont {Q.}~\bibnamefont
  {Huang}}, \bibinfo {author} {\bibfnamefont {E.}~\bibnamefont
  {Bourret-Courchesne}}, \bibinfo {author} {\bibfnamefont {G.}~\bibnamefont
  {Simutis}}, \bibinfo {author} {\bibfnamefont {R.}~\bibnamefont {Khasanov}},
  \bibinfo {author} {\bibfnamefont {D.-X.}\ \bibnamefont {Yao}},  \emph
  {et~al.},\ }\href {https://link.aps.org/doi/10.1103/PhysRevB.98.180402}
  {\bibfield  {journal} {\bibinfo  {journal} {Physical Review B}\ }\textbf
  {\bibinfo {volume} {98}},\ \bibinfo {pages} {180402} (\bibinfo {year}
  {2018})}\BibitemShut {NoStop}%
\bibitem [{\citenamefont {Torikachvili}\ \emph {et~al.}(2008)\citenamefont
  {Torikachvili}, \citenamefont {Bud’ko}, \citenamefont {Ni},\ and\
  \citenamefont {Canfield}}]{torikachvili2008pressure}%
  \BibitemOpen
  \bibfield  {author} {\bibinfo {author} {\bibfnamefont {M.~S.}\ \bibnamefont
  {Torikachvili}}, \bibinfo {author} {\bibfnamefont {S.~L.}\ \bibnamefont
  {Bud’ko}}, \bibinfo {author} {\bibfnamefont {N.}~\bibnamefont {Ni}}, \ and\
  \bibinfo {author} {\bibfnamefont {P.~C.}\ \bibnamefont {Canfield}},\ }\href
  {https://link.aps.org/doi/10.1103/PhysRevLett.101.057006} {\bibfield
  {journal} {\bibinfo  {journal} {Physical Review Letters}\ }\textbf {\bibinfo
  {volume} {101}},\ \bibinfo {pages} {057006} (\bibinfo {year}
  {2008})}\BibitemShut {NoStop}%
\bibitem [{\citenamefont {Okada}\ \emph {et~al.}(2008)\citenamefont {Okada},
  \citenamefont {Igawa}, \citenamefont {Takahashi}, \citenamefont {Kamihara},
  \citenamefont {Hirano}, \citenamefont {Hosono}, \citenamefont
  {Matsubayashi},\ and\ \citenamefont {Uwatoko}}]{okada2008superconductivity}%
  \BibitemOpen
  \bibfield  {author} {\bibinfo {author} {\bibfnamefont {H.}~\bibnamefont
  {Okada}}, \bibinfo {author} {\bibfnamefont {K.}~\bibnamefont {Igawa}},
  \bibinfo {author} {\bibfnamefont {H.}~\bibnamefont {Takahashi}}, \bibinfo
  {author} {\bibfnamefont {Y.}~\bibnamefont {Kamihara}}, \bibinfo {author}
  {\bibfnamefont {M.}~\bibnamefont {Hirano}}, \bibinfo {author} {\bibfnamefont
  {H.}~\bibnamefont {Hosono}}, \bibinfo {author} {\bibfnamefont
  {K.}~\bibnamefont {Matsubayashi}}, \ and\ \bibinfo {author} {\bibfnamefont
  {Y.}~\bibnamefont {Uwatoko}},\ }\href {\doibase
  https://doi.org/10.1143/jpsj.77.113712} {\bibfield  {journal} {\bibinfo
  {journal} {Journal of the Physical Society of Japan}\ }\textbf {\bibinfo
  {volume} {77}},\ \bibinfo {pages} {113712} (\bibinfo {year}
  {2008})}\BibitemShut {NoStop}%
\bibitem [{\citenamefont {Sun}\ \emph {et~al.}(2020)\citenamefont {Sun},
  \citenamefont {Li}, \citenamefont {Zhou}, \citenamefont {Yu}, \citenamefont
  {Frandsen}, \citenamefont {Wu}, \citenamefont {Xu}, \citenamefont {Jiang},
  \citenamefont {Huang}, \citenamefont {Bourret-Courchesne}, \citenamefont
  {Sun}, \citenamefont {Lynn}, \citenamefont {Birgeneau},\ and\ \citenamefont
  {Wang}}]{PhysRevB.101.205129}%
  \BibitemOpen
  \bibfield  {author} {\bibinfo {author} {\bibfnamefont {H.}~\bibnamefont
  {Sun}}, \bibinfo {author} {\bibfnamefont {X.}~\bibnamefont {Li}}, \bibinfo
  {author} {\bibfnamefont {Y.}~\bibnamefont {Zhou}}, \bibinfo {author}
  {\bibfnamefont {J.}~\bibnamefont {Yu}}, \bibinfo {author} {\bibfnamefont
  {B.~A.}\ \bibnamefont {Frandsen}}, \bibinfo {author} {\bibfnamefont
  {S.}~\bibnamefont {Wu}}, \bibinfo {author} {\bibfnamefont {Z.}~\bibnamefont
  {Xu}}, \bibinfo {author} {\bibfnamefont {S.}~\bibnamefont {Jiang}}, \bibinfo
  {author} {\bibfnamefont {Q.}~\bibnamefont {Huang}}, \bibinfo {author}
  {\bibfnamefont {E.}~\bibnamefont {Bourret-Courchesne}}, \bibinfo {author}
  {\bibfnamefont {L.}~\bibnamefont {Sun}}, \bibinfo {author} {\bibfnamefont
  {J.~W.}\ \bibnamefont {Lynn}}, \bibinfo {author} {\bibfnamefont {R.~J.}\
  \bibnamefont {Birgeneau}}, \ and\ \bibinfo {author} {\bibfnamefont
  {M.}~\bibnamefont {Wang}},\ }\href {\doibase 10.1103/PhysRevB.101.205129}
  {\bibfield  {journal} {\bibinfo  {journal} {Phys. Rev. B}\ }\textbf {\bibinfo
  {volume} {101}},\ \bibinfo {pages} {205129} (\bibinfo {year}
  {2020})}\BibitemShut {NoStop}%
\bibitem [{\citenamefont {Hirata}\ \emph {et~al.}(2015)\citenamefont {Hirata},
  \citenamefont {Maki}, \citenamefont {Yamaura}, \citenamefont {Yamauchi},\
  and\ \citenamefont {Ohgushi}}]{hirata2015effects}%
  \BibitemOpen
  \bibfield  {author} {\bibinfo {author} {\bibfnamefont {Y.}~\bibnamefont
  {Hirata}}, \bibinfo {author} {\bibfnamefont {S.}~\bibnamefont {Maki}},
  \bibinfo {author} {\bibfnamefont {J.-i.}\ \bibnamefont {Yamaura}}, \bibinfo
  {author} {\bibfnamefont {T.}~\bibnamefont {Yamauchi}}, \ and\ \bibinfo
  {author} {\bibfnamefont {K.}~\bibnamefont {Ohgushi}},\ }\href
  {https://link.aps.org/doi/10.1103/PhysRevB.92.205109} {\bibfield  {journal}
  {\bibinfo  {journal} {Physical Review B}\ }\textbf {\bibinfo {volume} {92}},\
  \bibinfo {pages} {205109} (\bibinfo {year} {2015})}\BibitemShut {NoStop}%
\bibitem [{\citenamefont {Stinn}\ \emph {et~al.}(2017)\citenamefont {Stinn},
  \citenamefont {Nose}, \citenamefont {Okabe},\ and\ \citenamefont
  {Allanore}}]{stinn2017experimentally}%
  \BibitemOpen
  \bibfield  {author} {\bibinfo {author} {\bibfnamefont {C.}~\bibnamefont
  {Stinn}}, \bibinfo {author} {\bibfnamefont {K.}~\bibnamefont {Nose}},
  \bibinfo {author} {\bibfnamefont {T.}~\bibnamefont {Okabe}}, \ and\ \bibinfo
  {author} {\bibfnamefont {A.}~\bibnamefont {Allanore}},\ }\href
  {https://doi.org/10.1007/s11663-017-1107-5} {\bibfield  {journal} {\bibinfo
  {journal} {Metallurgical and Materials Transactions B}\ }\textbf {\bibinfo
  {volume} {48}},\ \bibinfo {pages} {2922} (\bibinfo {year}
  {2017})}\BibitemShut {NoStop}%
\bibitem [{\citenamefont {Neubronner}\ \emph {et~al.}()\citenamefont
  {Neubronner}, \citenamefont {Bodmer}, \citenamefont {H{\"u}bner},
  \citenamefont {Kempa}, \citenamefont {Tsotsas}, \citenamefont {Eschner},
  \citenamefont {Kasparek}, \citenamefont {Ochs}, \citenamefont
  {M{\"u}ller-Steinhagen}, \citenamefont {Werner},\ and\ \citenamefont
  {Spitzner}}]{SpringerMaterialsAdditionalResources2010:sm_nlb_978-3-540-77877-6_26}%
  \BibitemOpen
  \bibfield  {author} {\bibinfo {author} {\bibfnamefont {M.}~\bibnamefont
  {Neubronner}}, \bibinfo {author} {\bibfnamefont {T.}~\bibnamefont {Bodmer}},
  \bibinfo {author} {\bibfnamefont {C.}~\bibnamefont {H{\"u}bner}}, \bibinfo
  {author} {\bibfnamefont {P.~B.}\ \bibnamefont {Kempa}}, \bibinfo {author}
  {\bibfnamefont {E.}~\bibnamefont {Tsotsas}}, \bibinfo {author} {\bibfnamefont
  {A.}~\bibnamefont {Eschner}}, \bibinfo {author} {\bibfnamefont
  {G.}~\bibnamefont {Kasparek}}, \bibinfo {author} {\bibfnamefont
  {F.}~\bibnamefont {Ochs}}, \bibinfo {author} {\bibfnamefont {H.}~\bibnamefont
  {M{\"u}ller-Steinhagen}}, \bibinfo {author} {\bibfnamefont {H.}~\bibnamefont
  {Werner}}, \ and\ \bibinfo {author} {\bibfnamefont {M.~H.}\ \bibnamefont
  {Spitzner}},\ }\href
  {https://materials.springer.com/lb/docs/sm_nlb_978-3-540-77877-6_26}
  {\enquote {\bibinfo {title} {D6 properties of solids and solid materials:
  Datasheet from {VDI-Buch}},}\ }\BibitemShut {NoStop}%
\bibitem [{\citenamefont
  {Rodr{\'\i}guez-Carvajal}(1993)}]{rodriguez1993recent}%
  \BibitemOpen
  \bibfield  {author} {\bibinfo {author} {\bibfnamefont {J.}~\bibnamefont
  {Rodr{\'\i}guez-Carvajal}},\ }\href
  {https://doi.org/10.1016/0921-4526(93)90108-I} {\bibfield  {journal}
  {\bibinfo  {journal} {Physica B: Condensed Matter}\ }\textbf {\bibinfo
  {volume} {192}},\ \bibinfo {pages} {55} (\bibinfo {year} {1993})}\BibitemShut
  {NoStop}%
\bibitem [{APE(2017)}]{APEX3}%
  \BibitemOpen
  \href@noop {} {\bibfield  {journal} {\bibinfo  {journal} {Bruker, APEX3
  v2018.1-0, Bruker AXS Inc., Madison, Wisconsin, USA}\ } (\bibinfo {year}
  {2017})}\BibitemShut {NoStop}%
\bibitem [{SAI(2017)}]{SAINT}%
  \BibitemOpen
  \href@noop {} {\bibfield  {journal} {\bibinfo  {journal} {SAINT(V8.30A),
  Bruker AXS Inc., Madison, Wisconsin, USA}\ } (\bibinfo {year}
  {2017})}\BibitemShut {NoStop}%
\bibitem [{\citenamefont {Krause}\ \emph {et~al.}(2015)\citenamefont {Krause},
  \citenamefont {Herbst-Irmer}, \citenamefont {Sheldrick},\ and\ \citenamefont
  {Stalke}}]{SADABS}%
  \BibitemOpen
  \bibfield  {author} {\bibinfo {author} {\bibfnamefont {L.}~\bibnamefont
  {Krause}}, \bibinfo {author} {\bibfnamefont {R.}~\bibnamefont
  {Herbst-Irmer}}, \bibinfo {author} {\bibfnamefont {G.~M.}\ \bibnamefont
  {Sheldrick}}, \ and\ \bibinfo {author} {\bibfnamefont {D.}~\bibnamefont
  {Stalke}},\ }\href {https://doi.org/10.1107/S1600576714022985} {\bibfield
  {journal} {\bibinfo  {journal} {Journal of Applied Crystallography}\ }\textbf
  {\bibinfo {volume} {48}},\ \bibinfo {pages} {3} (\bibinfo {year}
  {2015})}\BibitemShut {NoStop}%
\bibitem [{\citenamefont {Sheldrick}(2008)}]{SHELX}%
  \BibitemOpen
  \bibfield  {author} {\bibinfo {author} {\bibfnamefont {G.~M.}\ \bibnamefont
  {Sheldrick}},\ }\href {https://doi.org/10.1107/S0108767307043930} {\bibfield
  {journal} {\bibinfo  {journal} {Acta Crystallographica Section A: Foundations
  of Crystallography}\ }\textbf {\bibinfo {volume} {64}},\ \bibinfo {pages}
  {112} (\bibinfo {year} {2008})}\BibitemShut {NoStop}%
\bibitem [{\citenamefont {Farrugia}(1999)}]{WinGX}%
  \BibitemOpen
  \bibfield  {author} {\bibinfo {author} {\bibfnamefont {L.~J.}\ \bibnamefont
  {Farrugia}},\ }\href {https://doi.org/10.1107/S0021889899006020} {\bibfield
  {journal} {\bibinfo  {journal} {Journal of Applied Crystallography}\ }\textbf
  {\bibinfo {volume} {32}},\ \bibinfo {pages} {837} (\bibinfo {year}
  {1999})}\BibitemShut {NoStop}%
\bibitem [{\citenamefont {Swinnea}\ \emph {et~al.}(1982)\citenamefont
  {Swinnea}, \citenamefont {Eisman}, \citenamefont {Perng}, \citenamefont
  {Kimizuka},\ and\ \citenamefont {Steinfink}}]{swinnea1982crystal}%
  \BibitemOpen
  \bibfield  {author} {\bibinfo {author} {\bibfnamefont {J.~S.}\ \bibnamefont
  {Swinnea}}, \bibinfo {author} {\bibfnamefont {G.}~\bibnamefont {Eisman}},
  \bibinfo {author} {\bibfnamefont {T.}~\bibnamefont {Perng}}, \bibinfo
  {author} {\bibfnamefont {N.}~\bibnamefont {Kimizuka}}, \ and\ \bibinfo
  {author} {\bibfnamefont {H.}~\bibnamefont {Steinfink}},\ }\href
  {https://doi.org/10.1016/0022-4596(82)90040-8} {\bibfield  {journal}
  {\bibinfo  {journal} {Journal of Solid State Chemistry}\ }\textbf {\bibinfo
  {volume} {41}},\ \bibinfo {pages} {104} (\bibinfo {year} {1982})}\BibitemShut
  {NoStop}%
\bibitem [{\citenamefont {Marshall}\ \emph {et~al.}(2000)\citenamefont
  {Marshall}, \citenamefont {Nelmes}, \citenamefont {Loveday}, \citenamefont
  {Klotz}, \citenamefont {Besson}, \citenamefont {Hamel},\ and\ \citenamefont
  {Parise}}]{marshall2000high}%
  \BibitemOpen
  \bibfield  {author} {\bibinfo {author} {\bibfnamefont {W.}~\bibnamefont
  {Marshall}}, \bibinfo {author} {\bibfnamefont {R.}~\bibnamefont {Nelmes}},
  \bibinfo {author} {\bibfnamefont {J.}~\bibnamefont {Loveday}}, \bibinfo
  {author} {\bibfnamefont {S.}~\bibnamefont {Klotz}}, \bibinfo {author}
  {\bibfnamefont {J.}~\bibnamefont {Besson}}, \bibinfo {author} {\bibfnamefont
  {G.}~\bibnamefont {Hamel}}, \ and\ \bibinfo {author} {\bibfnamefont
  {J.}~\bibnamefont {Parise}},\ }\href {\doibase 10.1103/PhysRevB.61.11201}
  {\bibfield  {journal} {\bibinfo  {journal} {Physical Review B}\ }\textbf
  {\bibinfo {volume} {61}},\ \bibinfo {pages} {11201} (\bibinfo {year}
  {2000})}\BibitemShut {NoStop}%
\bibitem [{\citenamefont {Saparov}\ \emph {et~al.}(2011)\citenamefont
  {Saparov}, \citenamefont {Calder}, \citenamefont {Sipos}, \citenamefont
  {Cao}, \citenamefont {Chi}, \citenamefont {Singh}, \citenamefont
  {Christianson}, \citenamefont {Lumsden},\ and\ \citenamefont
  {Sefat}}]{Saparov}%
  \BibitemOpen
  \bibfield  {author} {\bibinfo {author} {\bibfnamefont {B.}~\bibnamefont
  {Saparov}}, \bibinfo {author} {\bibfnamefont {S.}~\bibnamefont {Calder}},
  \bibinfo {author} {\bibfnamefont {B.}~\bibnamefont {Sipos}}, \bibinfo
  {author} {\bibfnamefont {H.}~\bibnamefont {Cao}}, \bibinfo {author}
  {\bibfnamefont {S.}~\bibnamefont {Chi}}, \bibinfo {author} {\bibfnamefont
  {D.~J.}\ \bibnamefont {Singh}}, \bibinfo {author} {\bibfnamefont {A.~D.}\
  \bibnamefont {Christianson}}, \bibinfo {author} {\bibfnamefont {M.~D.}\
  \bibnamefont {Lumsden}}, \ and\ \bibinfo {author} {\bibfnamefont {A.~S.}\
  \bibnamefont {Sefat}},\ }\href {\doibase 10.1103/PhysRevB.84.245132}
  {\bibfield  {journal} {\bibinfo  {journal} {Physical Review B}\ }\textbf
  {\bibinfo {volume} {84}},\ \bibinfo {pages} {245132} (\bibinfo {year}
  {2011})}\BibitemShut {NoStop}%
\bibitem [{\citenamefont {Svitlyk}\ \emph {et~al.}(2013)\citenamefont
  {Svitlyk}, \citenamefont {Chernyshov}, \citenamefont {Pomjakushina},
  \citenamefont {Krzton-Maziopa}, \citenamefont {Conder}, \citenamefont
  {Pomjakushin}, \citenamefont {P{\"o}ttgen},\ and\ \citenamefont
  {Dmitriev}}]{Svitlyk}%
  \BibitemOpen
  \bibfield  {author} {\bibinfo {author} {\bibfnamefont {V.}~\bibnamefont
  {Svitlyk}}, \bibinfo {author} {\bibfnamefont {D.}~\bibnamefont {Chernyshov}},
  \bibinfo {author} {\bibfnamefont {E.}~\bibnamefont {Pomjakushina}}, \bibinfo
  {author} {\bibfnamefont {A.}~\bibnamefont {Krzton-Maziopa}}, \bibinfo
  {author} {\bibfnamefont {K.}~\bibnamefont {Conder}}, \bibinfo {author}
  {\bibfnamefont {V.}~\bibnamefont {Pomjakushin}}, \bibinfo {author}
  {\bibfnamefont {R.}~\bibnamefont {P{\"o}ttgen}}, \ and\ \bibinfo {author}
  {\bibfnamefont {V.}~\bibnamefont {Dmitriev}},\ }\href
  {https://doi.org/10.1088/0953-8984/25/31/315403} {\bibfield  {journal}
  {\bibinfo  {journal} {Journal of Physics: Condensed Matter}\ }\textbf
  {\bibinfo {volume} {25}},\ \bibinfo {pages} {315403} (\bibinfo {year}
  {2013})}\BibitemShut {NoStop}%
\bibitem [{\citenamefont {Johnston}(1996)}]{johnston1996antiferromagnetic}%
  \BibitemOpen
  \bibfield  {author} {\bibinfo {author} {\bibfnamefont {D.}~\bibnamefont
  {Johnston}},\ }\href {\doibase 10.1103/PhysRevB.54.13009} {\bibfield
  {journal} {\bibinfo  {journal} {Physical Review B}\ }\textbf {\bibinfo
  {volume} {54}},\ \bibinfo {pages} {13009} (\bibinfo {year}
  {1996})}\BibitemShut {NoStop}%
\bibitem [{\citenamefont {Tiwary}\ and\ \citenamefont
  {Vasudevan}(1997)}]{tiwary1997single}%
  \BibitemOpen
  \bibfield  {author} {\bibinfo {author} {\bibfnamefont {S.~K.}\ \bibnamefont
  {Tiwary}}\ and\ \bibinfo {author} {\bibfnamefont {S.}~\bibnamefont
  {Vasudevan}},\ }\href {https://doi.org/10.1016/S0038-1098(96)00617-5}
  {\bibfield  {journal} {\bibinfo  {journal} {Solid state communications}\
  }\textbf {\bibinfo {volume} {101}},\ \bibinfo {pages} {449} (\bibinfo {year}
  {1997})}\BibitemShut {NoStop}%
\bibitem [{\citenamefont {Zhang}\ \emph {et~al.}(2018)\citenamefont {Zhang},
  \citenamefont {Zhang}, \citenamefont {Ma}, \citenamefont {Wang},
  \citenamefont {Chu}, \citenamefont {Hu}, \citenamefont {Mu}, \citenamefont
  {Lu}, \citenamefont {Cai}, \citenamefont {Huang} \emph
  {et~al.}}]{zhang2018situ}%
  \BibitemOpen
  \bibfield  {author} {\bibinfo {author} {\bibfnamefont {X.}~\bibnamefont
  {Zhang}}, \bibinfo {author} {\bibfnamefont {H.}~\bibnamefont {Zhang}},
  \bibinfo {author} {\bibfnamefont {Y.}~\bibnamefont {Ma}}, \bibinfo {author}
  {\bibfnamefont {L.}~\bibnamefont {Wang}}, \bibinfo {author} {\bibfnamefont
  {J.}~\bibnamefont {Chu}}, \bibinfo {author} {\bibfnamefont {T.}~\bibnamefont
  {Hu}}, \bibinfo {author} {\bibfnamefont {G.}~\bibnamefont {Mu}}, \bibinfo
  {author} {\bibfnamefont {Y.}~\bibnamefont {Lu}}, \bibinfo {author}
  {\bibfnamefont {C.}~\bibnamefont {Cai}}, \bibinfo {author} {\bibfnamefont
  {F.}~\bibnamefont {Huang}},  \emph {et~al.},\ }\href
  {https://doi.org/10.1007/s11433-017-9192-4} {\bibfield  {journal} {\bibinfo
  {journal} {Science China Physics, Mechanics \& Astronomy}\ }\textbf {\bibinfo
  {volume} {61}},\ \bibinfo {pages} {077421} (\bibinfo {year}
  {2018})}\BibitemShut {NoStop}%
\bibitem [{\citenamefont {Bates}(1948)}]{honda}%
  \BibitemOpen
  \bibfield  {author} {\bibinfo {author} {\bibfnamefont {L.~F.}\ \bibnamefont
  {Bates}},\ }\href@noop {} {\emph {\bibinfo {title} {Modern Magnetism}}}\
  (\bibinfo  {publisher} {Cambrige at university press},\ \bibinfo {year}
  {1948})\BibitemShut {NoStop}%
\bibitem [{\citenamefont {Hirahara}\ and\ \citenamefont
  {Murakami}(1958)}]{hirahara1958magnetic}%
  \BibitemOpen
  \bibfield  {author} {\bibinfo {author} {\bibfnamefont {E.}~\bibnamefont
  {Hirahara}}\ and\ \bibinfo {author} {\bibfnamefont {M.}~\bibnamefont
  {Murakami}},\ }\href {https://doi.org/10.1016/0022-3697(58)90278-6}
  {\bibfield  {journal} {\bibinfo  {journal} {Journal of Physics and Chemistry
  of Solids}\ }\textbf {\bibinfo {volume} {7}},\ \bibinfo {pages} {281}
  (\bibinfo {year} {1958})}\BibitemShut {NoStop}%
\bibitem [{\citenamefont {Fogler}\ \emph {et~al.}(2004)\citenamefont {Fogler},
  \citenamefont {Teber},\ and\ \citenamefont
  {Shklovskii}}]{fogler2004variable}%
  \BibitemOpen
  \bibfield  {author} {\bibinfo {author} {\bibfnamefont {M.}~\bibnamefont
  {Fogler}}, \bibinfo {author} {\bibfnamefont {S.}~\bibnamefont {Teber}}, \
  and\ \bibinfo {author} {\bibfnamefont {B.}~\bibnamefont {Shklovskii}},\
  }\href {\doibase 10.1103/PhysRevB.69.035413} {\bibfield  {journal} {\bibinfo
  {journal} {Physical Review B}\ }\textbf {\bibinfo {volume} {69}},\ \bibinfo
  {pages} {035413} (\bibinfo {year} {2004})}\BibitemShut {NoStop}%
\end{thebibliography}%
\bibliographystyle{apsrev4-1}


\appendix

\section{Single crystal x-ray diffraction}
\label{scxr}

\begin{table*}[h]
\caption{Details on data collection and structure refinement of BaFe$_{2+\delta}$S$_3$ as determined from single-crystal x-ray diffraction as a function of the nominal Fe content $\delta$. The lattice parameters were obtained from the Rietveld refinement of the powder x-ray diffraction data.} 
\centering 
\begin{tabular}{l c c c c c} 
\hline\hline 
nominal $\delta$ & 0 & 0.05 (re-melted) & 0.05 & 0.1 & 0.2 \\ 
[0.5ex] 
\hline 
Crystal data \\
\textit{Temperature} (K)& 295 & 295 & 295 & 295 & 295\\
\textit{Space group} & \textit{Cmcm} & \textit{Cmcm} & \textit{Cmcm} & \textit{Cmcm} & \textit{Cmcm}\\
\textit{a} (\AA) & 8.7759(2) & 8.7742(4) & 8.7762(3) & 8.7797(4) & 8.7765(3)\\  
\textit{b} (\AA) & 11.2177(3) & 11.2137(5) & 11.2151(4) & 11.2211(5) & 11.2199(4)\\
\textit{c} (\AA) & 5.2823(1) & 5.2793(2) & 5.2849(2) & 5.2850(2) & 5.2841(2)\\
\textit{Z}  & 4 & 4 & 4 & 4 & 4\\
\textit{M$_{r}$} & 345.17 & 345.17 & 345.17 & 345.17 & 345.17 \\
\textit{$\rho_{calc}$} (g$\cdot$cm$^{-3}$) & 4.408 & 4.412 & 4.408 & 4.406 & 4.407\\
\textit{$\mu$} (mm$^{-1}$) & 14.014 & 14.026 & 14.016 & 14.010 & 14.011\\
[1ex] 
Data collection \\
\textit{2 $\theta_{max}$} ($^\circ$) & 62.938 & 62.968 & 62.932 & 62.922 & 54.95\\
\textit{Absorption correction}  & Multi-Scan & Multi-Scan & Multi-Scan & Multi-Scan & Multi-Scan\\
\textit{T$_{min}$} & 0.5527 & 0.6714 & 0.6105 & 0.4501 & 0.4644\\
\textit{T$_{max}$} & 0.7462 & 0.7462 & 0.7462 & 0.7462 & 0.7462\\
\textit{N$_{measured}$} & 5269 & 6017 & 4441 & 4482 & 2295\\
\textit{N$_{independent}$} & 486 & 486 & 489 & 505 & 354\\
\textit{R$_{int}$} (\%) & 2.97 & 3.56 & 1.99 & 2.16 & 2.5\\
[1ex] 
Refinement \\
\textit{N$_{parameters}$} & 20 & 20 & 20 & 20 & 20\\
\textit{R$_{1} > 4\sigma$} (\%) & 1.69 & 1.48 & 1.11 & 1.14 & 1.27\\
\textit{R$_{1}$ all} (\%) & 2.33 & 2.07 & 1.36 & 1.28 & 1.44\\
\textit{wR$_{2} > 4\sigma$} (\%) & 3.12 & 2.83 & 2.29 & 2.52 & 2.60\\
\textit{wR$_{2}$ all} (\%) & 3.27 & 2.98 & 2.34 & 2.56 & 2.65\\
\textit{G.O.F} & 1.137 & 1.091 & 1.055 & 1.105 & 1.086\\
\textit{$\Delta\rho_{min}$} (e$\cdot$A$^{-3}$)& -0.756 & -0.797 & -0.371 & -0.670 & -0.372\\
\textit{$\Delta\rho_{max}$} (e$\cdot$A$^{-3}$) & 0.949 & 0.942 & 0.719 & 1.001 & 1.002\\
\textit{weight w (a,b)} & 0.0138 & 0.0136 & 0.0108 & 0.0094
& 0.0114 \\
& 1.0487 & 0.4001 & 0.5927 & 0.9146 & 0.2287 \\
[1ex] 
\hline \hline 
\end{tabular}
\label{table:DataCollection} 
\footnotesize{\begin{small} 
 
\textit{T$_{min}$} = minimum transmission, \textit{T$_{max}$} = maximum transmission, \textit{$R_{1}=\sum\parallel F_{0}\mid-\mid F_{c}\parallel/\sum \mid F_{0} \mid$}, \textit{$wR_{2}=\lbrace \sum [w(F_{0}^{2}-F_{c}^{2})^{2}]/ \sum [w(F_{0}^{2})^{2}]\rbrace ^{1/2}$}, \textit{$w=1/[(\sigma^{2}(F_{0}^{2}))+(aP)^{2}+bP]$} where \textit{$P=[2F_{c}^{2}+max(F_{0}^{2},0)]/3$}
\end{small}}
\end{table*}

\begin{table*}[h]
\caption{Atomic coordinates and equivalent isotropic displacement parameters of BaFe$_{2+\delta}$S$_3$ single crystals at 295 K as a function of the nominal Fe content $\delta$.} 
\centering 
\begin{tabular}{c c c c c c} 
\hline\hline 
nominal $\delta$ & 0 & 0.05 (re-melted) & 0.05 & 0.1 & 0.2 \\ [0.5ex] 
\hline 
Ba \\
\textit{x} & 0.5 & 0.5 & 0.5 & 0.5 & 0.5\\ 
\textit{y} & 0.18616(3) & 0.18623(3) & 0.18631(2) & 0.18633(2) & 0.18634(3)  \\
\textit{z} & 0.25 & 0.25 & 0.25 & 0.25 & 0.25  \\
\textit{U$_{eq}$} (\AA$^2$) & 0.01878(8) & 0.01893(8) & 0.01819(6) & 0.01881(6) & 0.01842(9) \\
Fe \\
\textit{x} & 0.34631(5) & 0.34636(4) & 0.34628(3) & 0.34630(3) & 0.34629(4)\\ 
\textit{y} & 0.5 & 0.5 & 0.5 & 0.5 & 0.5\\
\textit{z} & 0 & 0 & 0 & 0 & 0\\
\textit{U$_{eq}$} (\AA$^2$) & 0.01119(10) & 0.01155(9) & 0.01086(7) & 0.01148(7) & 0.01122(11)\\ [1ex] 
S$_{1}$  \\
\textit{x} &  0.5 & 0.5 & 0.5 & 0.5 & 0.5\\ 
\textit{y} &  0.61574(9) & 0.61580(8) & 0.61576(6) & 0.61575 & 0.61553(9)\\
\textit{z} &  0.25 & 0.25 & 0.25 & 0.25 & 0.25 \\
\textit{U$_{eq}$} (\AA$^2$) & 0.01135(20) & 0.01156(19) & 0.01114(13) & 0.01170(12) & 0.01151(20)\\ [1ex] 
S$_{2}$  \\
\textit{x} & 0.20756(11) & 0.20748(8) & 0.20749(6) & 0.20757(7) & 0.20746(9)\\ 
\textit{y} & 0.37843(8) & 0.37830(7) & 0.37831(5) & 0.37825(6) & 0.37836(8)\\
\textit{z} &  0.25 & 0.25 & 0.25 & 0.25 & 0.25 \\
\textit{U$_{eq}$} (\AA$^2$) & 0.01946(18) & 0.01950(16) & 0.01883(11) & 0.01940(11) & 0.01937(18)\\ [1ex] 
\hline \hline 
\end{tabular}
\label{table:Coordinate} 
\end{table*}

\begin{table*}[h]
\caption{Anisotropic displacement parameters of BaFe$_{2+\delta}$S$_3$ single crystals at 295 K as a function of the nominal Fe content $\delta$. } 
\centering 
\begin{tabular}{c c c c c c} 
\hline\hline 
nominal $\delta$ & 0 & 0.05 (re-melted) & 0.05 & 0.1 & 0.2  \\ [0.5ex] 
\hline 
Ba \\
\textit{U$_{11}$} (\AA$^2$) & 0.02021(17) & 0.01964(12) & 0.01934(9) & 0.01975(9) & 0.01835(15)\\  
\textit{U$_{22}$} (\AA$^2$) & 0.02223(14) & 0.02206(13) & 0.02123(9) & 0.02109(9) & 0.02160(16)\\
\textit{U$_{33}$} (\AA$^2$) & 0.01390(11) & 0.01508(15) & 0.01401(10) & 0.01560(9) & 0.01529(13)\\
Fe \\
\textit{U$_{11}$} (\AA$^2$) & 0.01072(24) & 0.01028(16) & 0.01046(12) & 0.01081(13) & 0.00958(21)\\
\textit{U$_{22}$} (\AA$^2$) & 0.01465(20) & 0.01509(18) & 0.01433(12) & 0.01409(13) & 0.01511(24)\\
\textit{U$_{33}$} (\AA$^2$) & 0.00821(17) & 0.00929(22) & 0.00779(14) & 0.00953(13) & 0.00899(17)\\
\textit{U$_{23}$} (\AA$^2$) & 0.00004(14) & 0.00030(15) & 0.00006(10) & 0.00016(10) & 0.00003(14)\\
S$_{1}$ \\
\textit{U$_{11}$} (\AA$^2$) & 0.01295(56) & 0.01197(37) & 0.01196(27) & 0.01245(28) & 0.01076(46)\\
\textit{U$_{22}$} (\AA$^2$) & 0.01101(44) & 0.01177(40) & 0.01154(26) & 0.01146(27) & 0.01248(52)\\
\textit{U$_{33}$} (\AA$^2$) & 0.01010(38) & 0.01096(51) & 0.00992(33) & 0.01118(29) & 0.01128(40)\\
S$_{2}$ \\
\textit{U$_{11}$} (\AA$^2$) & 0.01994(47) & 0.01908(31) & 0.01897(23) & 0.01960(24) & 0.01831(38)\\
\textit{U$_{22}$} (\AA$^2$) & 0.02679(41) & 0.02655(37) & 0.02577(25) & 0.02560(26) & 0.02657(45)\\
\textit{U$_{33}$} (\AA$^2$) & 0.01165(29) & 0.01388(41) & 0.01175(25) & 0.01301(23) & 0.01322(30)\\
\textit{U$_{12}$} (\AA$^2$) &-0.01233(37) &-0.01230(30) &-0.01198(21) &-0.01212(22) &-0.01256(18)\\ [1ex]
\hline \hline 
\end{tabular}
\label{table:Displacement} 
\end{table*}

\vspace{6cm}
\color{black}
\noindent\rule[0.25\baselineskip]{\textwidth}{1pt}

\end{document}